\documentclass[a4paper,11pt]{article}
\usepackage{jinstpub} 
\usepackage{lineno}
\usepackage{siunitx}
\nolinenumbers

\title{The Use of O$_2$ in Gas Mixtures for Drift Chambers}

\author[a]{A.~M.~Baldini,}
\author[b]{L.~Bianco,}
\author[a]{H.~Benmansour,}
\author[c]{G.~Cavoto,} 
\author[b,1]{F.~Cei,\note{Corresponding author.}}  
\author[a]{M.~Chiappini,} 
\author[e]{A.~Corvaglia,} 
\author[g]{M.~Francesconi,}
\author[c]{E.~Gabbrielli,}
\author[a]{L.~Galli,} 
\author[a]{G.~Gallucci,}
\author[e]{F.~Grancagnolo,} 
\author[b]{E.~G.~Grandoni,}
\author[a]{M.~Grassi,}
\author[b]{F.~Leonetti,}
\author[b]{D.~Nicol\'{o},} 
\author[f]{M.~Panareo,}
\author[d]{D.~Pasciuto,}
\author[b,h]{A.~Papa,}
\author[d]{F.~Renga,} 
\author[c]{S.~Scarpellini,}
\author[b]{A.~Venturini,}
\author[c]{C.~Voena.}

\affiliation[a]{INFN Sezione di Pisa \\Largo Bruno Pontecorvo, 3, 56127, Pisa, Italy}
\affiliation[b]{University of Pisa, Department of Physics and INFN Sezione di Pisa \\Largo Bruno Pontecorvo, 3, 56127, Pisa, Italy}
\affiliation[c]{University of Rome, La Sapienza Department of Physics and INFN Sezione di Roma, 
\\Piazzale A. Moro, 2, 00185, Rome, Italy}
\affiliation[d]{INFN Sezione di Roma, \\Piazzale A. Moro, 2, 00185, Rome, Italy}
\affiliation[e]{INFN Sezione di Lecce, \\Via per Arnesano, 73100, Lecce, Italy}
\affiliation[f]{University of Salento, Department of Mathematics and Physics and INFN Sezione di Lecce, \\Via per Arnesano, 73100, Lecce, Italy}
\affiliation[g]{INFN Sezione di Napoli, Dipartimento di Fisica dell`Universit\`a', Via Cintia, 80126 Napoli, Italy}
\affiliation[h]{Paul Scherrer Institut PSI, 5232 Villigen, Switzerland}

\emailAdd{fabrizio.cei@unipi.it}

\abstract{The use of Oxygen in gas mixtures for drift chambers is highly discouraged because Oxygen, being strongly electronegative, is generally believed to lead, even in very small quantities, to extremely reduced color{red} drift electron survival probability, thus preventing the detector's operation.

The drift chamber of the MEG II experiment at PSI has been operating for several years with a gas mixture that mainly contains He:Isobutane in relative proportions of 90:10 \% by molar concentration, in addition to 1.5\% Isopropanol and 0.5\% Oxygen. Oxygen and Isopropanol are essential for the proper functioning of the chamber. The electron attachment in the mixture used has proven negligible for the proper operation of the chamber and agrees well with the Garfield++ simulation after correctly accounting for the three-body attachment simulation.

However, it is important to note that, in the case of ternary or quaternary mixtures (as in the case of MEG II), it is not possible to provide Garfield++ with just the ingredients of the mixture used to obtain correct attachment values. Instead, the code needs to be properly modified and recompiled.
}


\arxivnumber{} 

\begin{document}

\maketitle

\section{Introduction}
Since their introduction in the early 1970s, drift chambers played a leading role in tracking applications for both collider and fixed-target experiments. With the advent of solid state strip and pixel detectors, the use of such detectors in very high energy and high radiation environments declined. However, drift chambers are still the technology of choice at lower energies, where the tracking resolutions are limited by the multiple Coulomb scattering in the detector material. Despite the introduction of very thin silicon pixel detectors like HV-MAPS, the extremely low material budget of gaseous chambers still makes their performance unmatched. Moreover, their relatively low cost makes them advantageous when large volumes need to be covered.

For these reasons, in the last decades drift chambers have been chosen for central tracking in the $\phi$-Factory and B-Factory experiments (KLOE~\cite{Calcaterra:1995mb}, BaBar~\cite{BaBarDriftChamber:1998kub}, Belle~\cite{Uno:1996gq} and Belle-II~\cite{Taniguchi:2017not}, BES-III~\cite{BESIII}) and in several low-energy applications like some muon decay experiments (TWIST~\cite{Henderson:2004zz}, MEG~\cite{Hildebrandt:2010zz} and MEG II~\cite{Afanaciev2024}, COMET~\cite{Sato:2024zbm}). The MEG II cylindrical drift chamber, in particular, marked a significant evolution in the construction technology of these detectors, allowing large volumes to be instrumented with high granularity, and paved the way for their use at future $e^+e^-$ colliders like FCC-ee, as proposed for the IDEA detector~\cite{Tassielli:2021rjk}.

When having a low material budget is critical, drift chambers are filled with extremely light, helium-based gas mixtures. A very common choice is He:Isobutane in relative proportions of 90:10~\% by molar concentration, which is proved to provide good stability at high gain (exceeding \SI{2e5}{} in the MEG II drift chamber). However, when exposed to high irradiation, chambers operated with this mixture tend to develop electron field emission from the electrodes, where polymers originating from Isobutane can deposit and form a thin insulating layer (Malter effect). It results in anomalously large currents and possibly electrical breakdowns. It was observed anyway that the addition of low concentrations of water or isopropyl alcohol vapors can restore the correct operation of the detector.

In the case of the MEG II Drift Chamber, deposits were observed in large portions of wires after large discharges happened during the commissioning of the detector. After testing several additives, a normal operation could be restored with the addition of isopropyl alcohol and a significant concentration of Oxygen, which is usually regarded as an unwanted contaminant because of the tendency of drifting electrons to attach to it, resulting in a suppression of the induced signals. 

In this paper we present experimental and simulation studies of the impact of Oxygen concentrations at a few permille level on some electron drift properties of the He:Isobutane 90:10 mixture. Our results support the semi-quantitative observation made in MEG II that similar concentrations of Oxygen can be used in drift chambers with small enough cells without compromising the overall performance of the detector. They will serve as a basis for the development of strategies to stabilize the operation of present and upcoming drift chambers, and could be critical in particular for the successful deployment of the drift chamber of the IDEA detector at FCC-ee.

\section{The MEG II Drift Chamber and Its Operation}

\subsection{The MEG II Drift Chamber}

\begin{figure}[htbp]
\centering
\includegraphics[width=.9\textwidth]{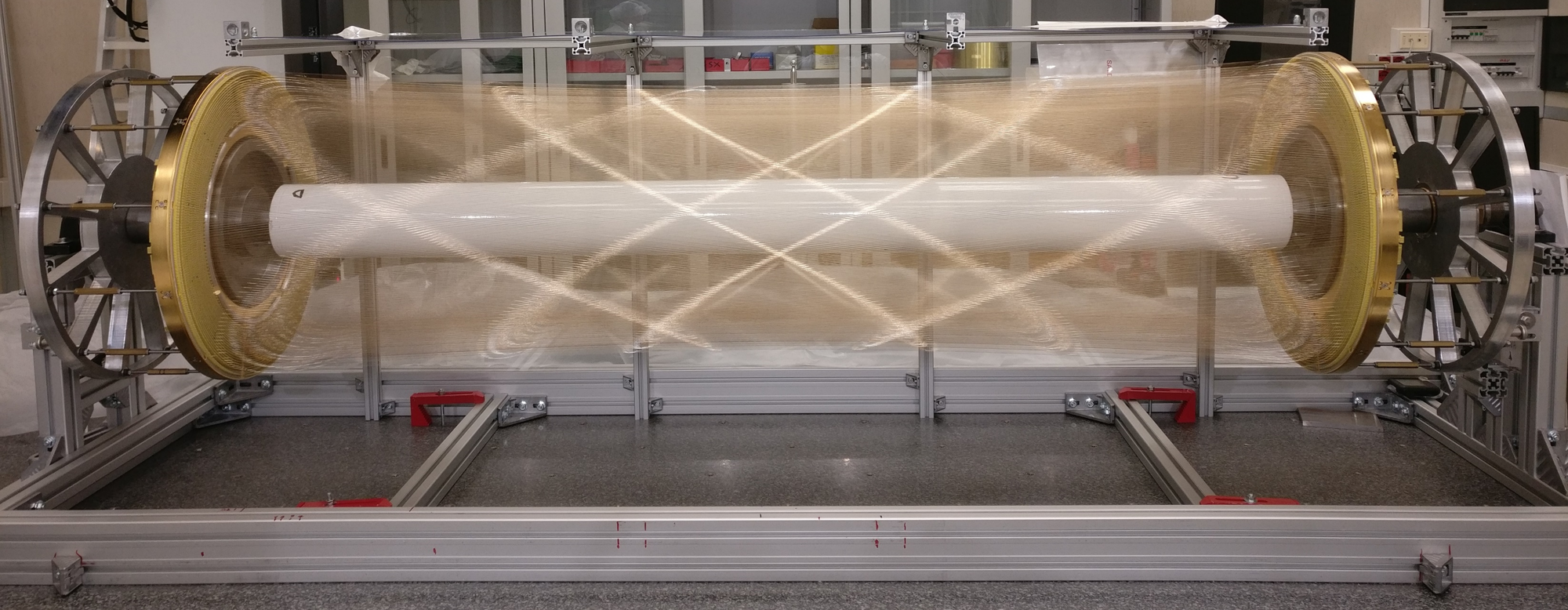}
\caption{The internal structures of the MEG II drift chamber\label{fig:CDCH}}
\end{figure}
The drift chamber of the MEG II experiment (described in more detail in the detector article \cite{Afanaciev2024} and the chamber performance article \cite{Baldini2024}), shown in Fig.~\ref{fig:CDCH}, allows the reconstruction of positron tracks with a momentum of about 50 MeV/c in a non-uniform cylindrical magnetic field with a central intensity of 1.27 T, decreasing along the central axis of the chamber. It is a unique volume detector, consisting of a 1.93 m long and 35 cm external diameter cylinder filled with a gas mixture of Helium:Isobutane (C$_4$H$_{10}$) 90:10, with small percentages of Isopropyl Alcohol (\SI{1.5}{\percent}) and Oxygen (\SI{0.5}{\percent}) added to stabilize the chamber's performance, as we will see shortly.

The volume is equipped with nine concentric layers of 192 gold-plated tungsten sense wires each, totaling 1728 wires, arranged in a stereo configuration with two views and approximately 10,000 silver-plated aluminum cathode and guard wires. The sense wires (20 micron diameter) collect the signals from the drift electrons, while the cathode and guard wires (40 and 50 micron diameter) form nearly squared drift cells and define the electric field within and at the boundaries of the sensitive volume; the cell dimensions range from 5.8 mm to 7.5 mm at the center and from 6.7 mm to 8.7 mm at the end-plates.

The High Voltage (HV) operating point was set in the range of 1400-1480 V, with the outermost layer at the highest voltage, to achieve a gas gain of approximately $2.5 \times 10^5$. The HV can be set for groups of eight cells in 10 V steps which enables to account for differences in cell size due to their radial position within the chamber.

Approximately $4 \times 10^7$ positive muons per second are typically stopped in a target at the chamber's center, exposing it to an intense flux of positrons produced by muon decay into a positron and two neutrinos. The normal current level observed for the eight consecutive anode wires (one "sector") powered by a single HV channel is around 10–20 $\mu A$. During initial operations with the pure He:Isobutane (90:10) mixture, an abrupt increase in current up to 400 $\mu A$ was observed for some HV channels.

A deep investigation revealed the formation of corona-like discharges along some wires. Several additives to the gas mixture were tested to restore normal detector operation. The reduction in high currents was achieved with an Oxygen level up to \SI{2}{\percent}, then gradually lowered to minimize electron attachment effects. 
\begin{figure}[htbp]
\centering
\includegraphics[width=.7\textwidth]{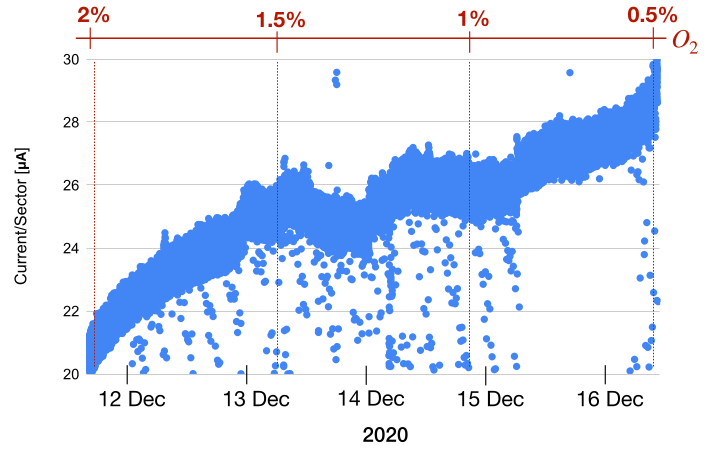}
\caption{Currents in one sector of the drift chamber when reducing Oxygen content from $2~\% \rightarrow 1~\% \rightarrow 0.5~\% \rightarrow 0.1~\%$. The abrupt reductions of the current were
due to instabilities of the PSI particle beam \label{fig:Current vs Oxygen}}
\end{figure}
The effect of Oxygen on the current per sector is shown in Fig.~\ref{fig:Current vs Oxygen}: current is lowered due to Oxygen capture and some loss of gas gain. In addition, a small percentage of Isopropyl Alcohol (also called Isopropanol, C$_3$H$_{7}$OH) was added to maintain a stable current level from the beginning of the stabilization procedure.

\subsection{Gas system upgrade for the addition of Oxygen and Isopropyl Alcohol}

The gas distribution system of the MEG II experiment is described in details in~\cite{Baldini:2018ing}. The system was designed to supply two separate volumes: the drift chamber, with a mixture of Helium and Isobutane, and the volume around the muon stopping target, with pure Helium. However, it was also prepared to include possible additives to the nominal gases. One additional mass flow controller was indeed included in each of the two supply lines. Thanks to this feature, the setup could be adapted to provide Alcohol and Oxygen to the nominal drift chamber mixture, with relatively small modifications of piping.

Oxygen is included in the mixture through a mass flow controller having a full scale of \SI{100}{sccm}\footnote{\lq\lq sccm\rq\rq stands for \lq\lq standard cubic cm per minute\rq\rq, where standard means measured in STP conditions.} for Nitrogen, corresponding to approximately \SI{99}{sccm} for Oxygen. Given a typical total gas flow of about \SI{500}{sccm}, the required flow for a \SI{0.5}{\percent} concentration of Oxygen is \SI{2.5}{sccm}, which is near the lower limit of the dynamic range of the mass flow controller (corresponding to \SI{2}{\percent} of the full scale). The Oxygen concentration is validated and monitored through a percent Oxygen analyzer with \SI{0.01}{\percent} precision periodically calibrated against a precision mixture of Nitrogen:Oxygen 80:20.

\begin{figure}[htbp]
\centering
\includegraphics[width=0.8\textwidth]{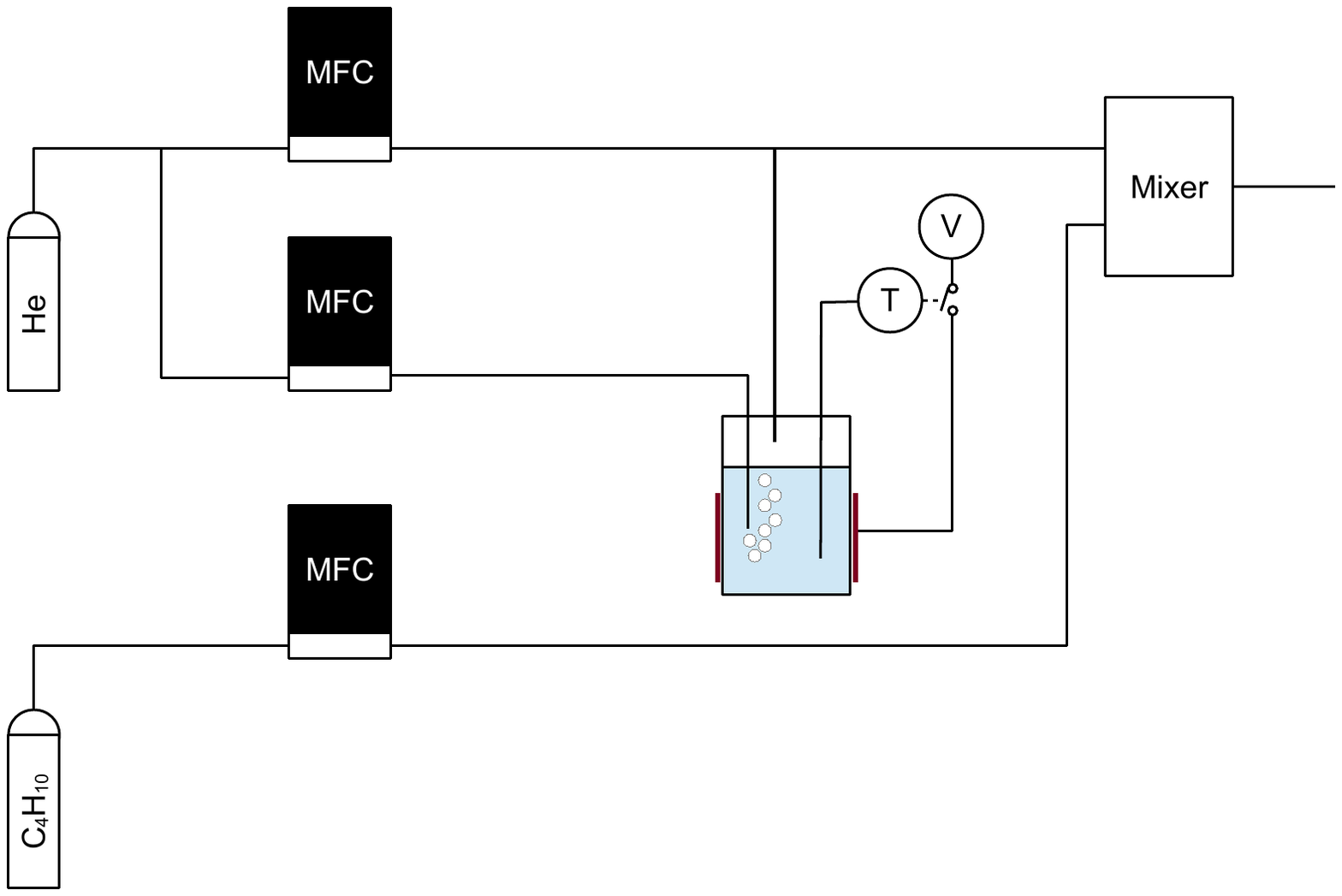}
\caption{Simplified schematic of the setup for the inclusion of Alcohol in the gas mixture, with bottles of Helium and Isobutane, mass flow controllers (MFC) to regulate the flows, and the bubbler filled with alcohol (in light blue) with its temperature probe (T) controlling the supply voltage (V) to the heating pads (in red). Gases are finally mixed in the mixing buffer as described in~\cite{Baldini:2018ing}. An additional MFC (not shown in this schematic) is used to add Oxygen to the Helium line before it is mixed with Isobutane. \label{fig:bubbler}}
\end{figure}

Isopropyl Alcohol is added to the mixture by sampling a fraction of the Helium (before mixing it with Isobutane) and flowing it through a custom-designed stainless-steel bubbler, where Alcohol is kept at a constant temperature of \SI{35}{\celsius} (measured by a Pt100 probe immersed in the liquid) by silicone heating pads connected to a thermostat. Helium exiting from the bubbler is saturated of Alcohol, with a partial pressure given by the semi-empirical Antoine~\cite{Reid} equation:
\begin{equation}
    \log_{10}P_{\mathrm{sat}} = A - \frac{B}{C + T}
\end{equation}
where $T = \SI{35}{\celsius}$ is the saturation temperature, and A = 8.00308, B = 1505.52, C = 211.6 for Isopropyl Alcohol~\cite{AntoineIsop} when the pressure is in mmHg. The concentration of Alcohol in the saturated flow is consequently given by $P_{\mathrm{sat}}/P_{\mathrm{atm}}$, where $P_{\mathrm{atm}}$ is the atmospheric pressure. Finally, selecting the fraction of the total Helium flow sent through the bubbler allows to regulate the Alcohol concentration in the total mixture. The actual concentration of Alcohol in the mixture cannot be measured with simple analyzers. Consequently, it is validated by measuring the daily consumption of Alcohol through a scale over which the bubbler is installed. Fig.~\ref{fig:bubbler} shows a simplified schematic of the setup. A system of valves (not shown in the schematic) allows to refill the bubbler without removing it from the circuit, with only a few minute interruption of flows, during which the chamber volume is kept isolated to prevent undesired contaminations.

The flows of the different gases are defined in such a way that the following ratios, in terms of molar concentrations, are satisfied: Helium:Isobutane 90:10; Alchool:Helium+Isobutane 1.5:98.5; Oxygen:Helium+Isobutane+Alcohol 0.5:99.5. The final composition of the mixture in termsof molar is approximately Helium:Isobutane:Alcohol:Oxygen 88.2:9.8:1.49:0.5. The concentration of Alcohol was changed over the MEG II data taking up to a maximum of \SI{1.9}{\percent} in order to improve the stability of the detector.

The concentration of Oxygen was measured to be sufficiently stable, with absolute variations within a few \SI{0.01}{\percent} over the annual MEG II data taking periods, without requiring any particular action. Keeping constant the concentration of Alcohol was proved to be much more difficult. It is well known that including vapors in a mixture is subject to uncertainties related to the actual saturation temperature, which can depend not only on the liquid temperature, but also on the temperature of the pipes the saturated gas flows through. Moreover, large and rapid excursions of the ambient temperature, typically happening overnight, can induce condensation of vapors inside the pipes if their thermal isolation is not sufficient, producing a temporary reduction of the Alcohol concentration inside the chamber. However, with incremental corrections of piping and isolation, we could finally get a satisfactory stability of Alcohol concentration, with only a few occurrences per year of significant drops, which are clearly visible as an increase of spark-related currents in the drift chamber. It is worth noting, however, that the relatively good stability in ambient temperature inside the PSI experimental hall (mostly within a few degrees during beam periods) played a crucial role in making this achievement possible.

\subsection{Chamber's Operation and Drift Electron Attachment}

\begin{figure}[htbp]
\centering
\includegraphics[width=.5\textwidth]{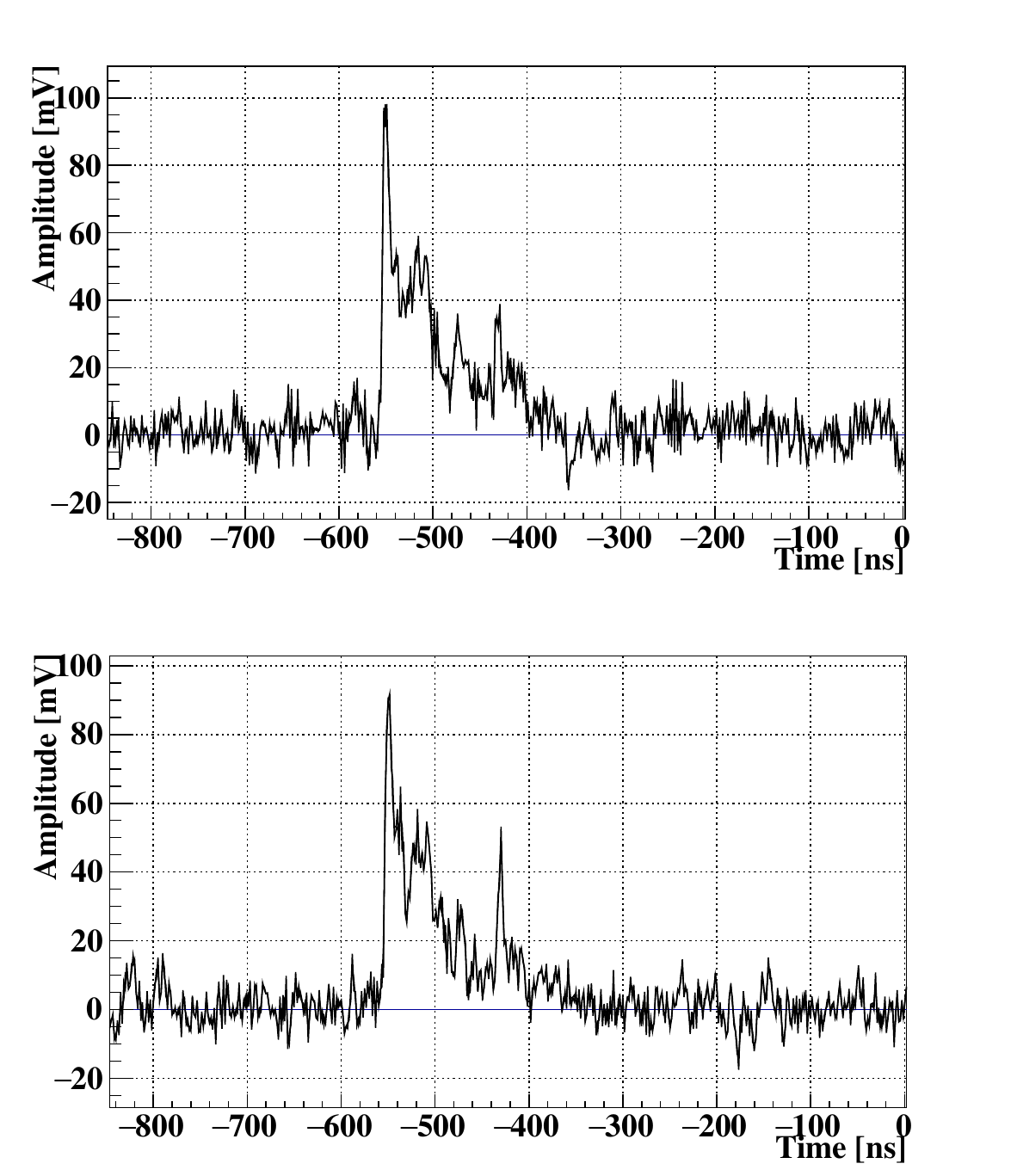}
\caption{An example of the waveforms recorded from both sides of one anode \label{fig:CDCHwfm}}
\end{figure}
The signal from each anode of the chamber is recorded by an analog sampler at a frequency of 1.2 Giga samples per second (GSPS) \cite{Galli2019}. The track reconstruction from the anode signals, an example of which is shown in Fig.~\ref{fig:CDCHwfm}, is a complex procedure described in detail in \cite{Baldini2024}.
\begin{figure}[htbp]
\centering
\includegraphics[width=.47\textwidth]{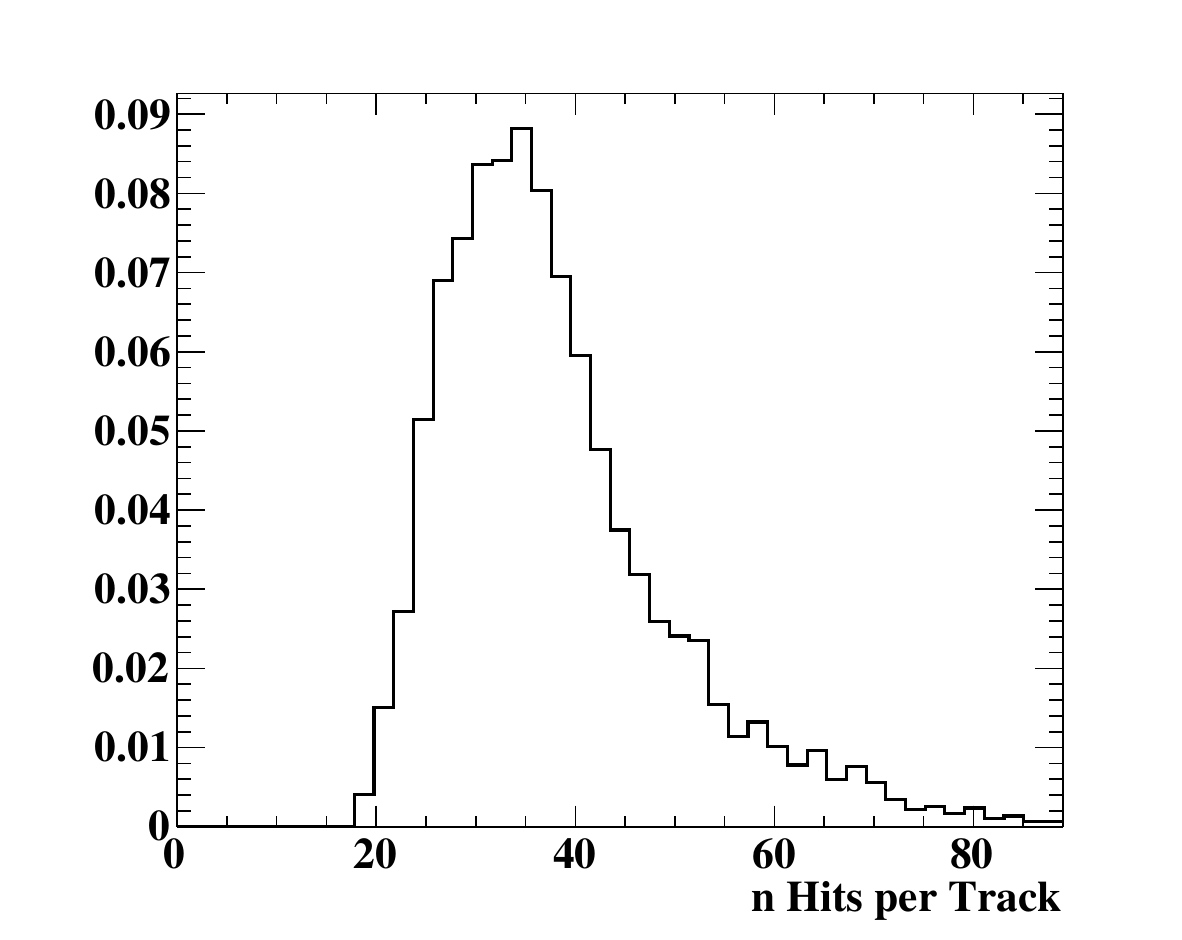}
\qquad
\includegraphics[width=.47\textwidth]
{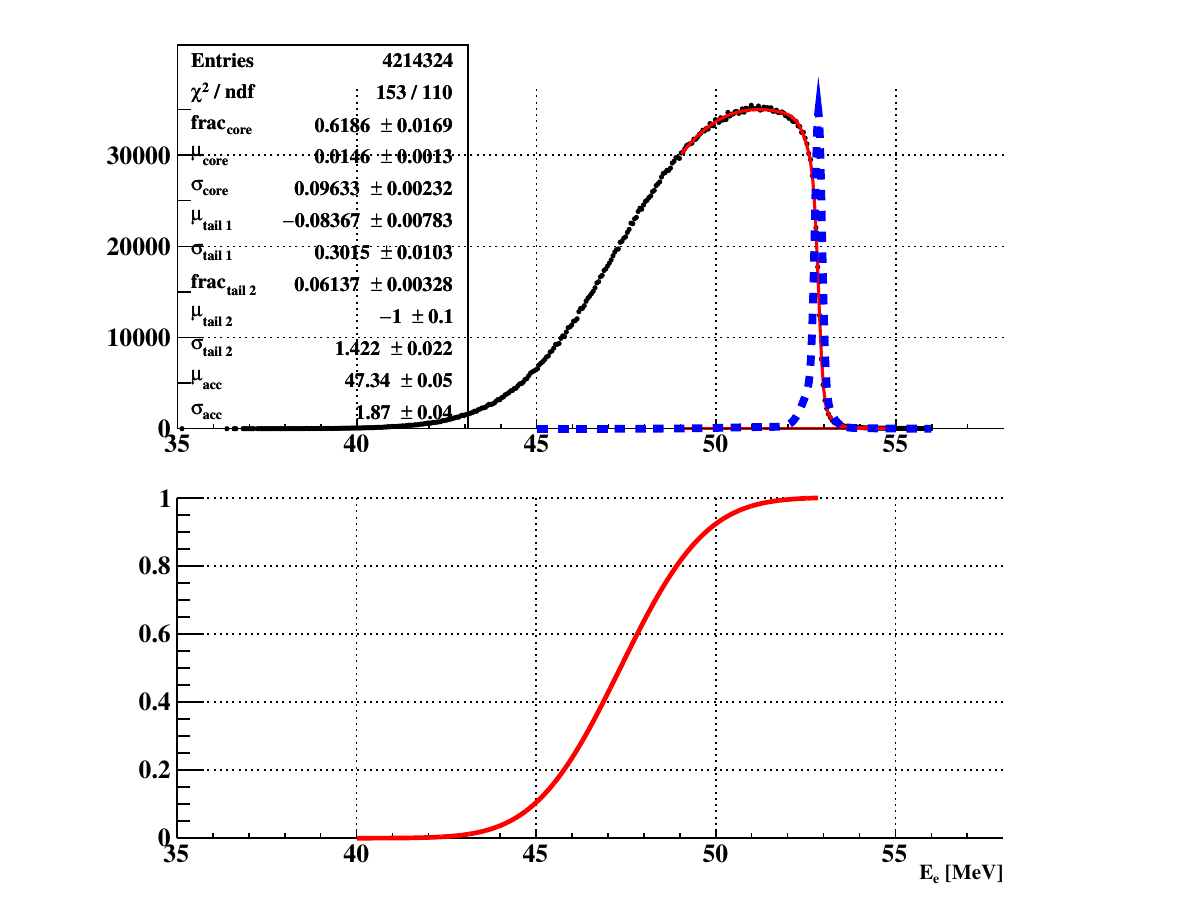}
\caption{Number of hits reconstructed per track (left) and positron momentum resolution (right)  obtained by fitting the Michel spectrum \label{fig:tracking}}
\end{figure}
The average number of reconstructed signals for a Michel positron is about 40 (see Fig.~\ref{fig:tracking}, left), and the achievable momentum resolution from the Michel positron  distribution, shown in Fig.~\ref{fig:tracking} (right), is approximately 90 keV/c. The reconstruction efficiency of positrons at a muon stopping rate $R_{\mu} = \SI{4e7}{}$ per second is \SI{70}{\percent}. The reconstruction efficiency and momentum resolution are in agreement with the initial estimates made during the design phase of the experiment, where the assumed mixture was Helium:Isobutane 90:10 without additives. The addition of \SI{0.5}{\percent} Oxygen and \SI{1.5}{\percent} Isopropyl Alcohol does not therefore seem to have a significant effect on the track reconstruction performance of the chamber.

\begin{figure}[htbp]
\centering
\includegraphics[width=.49\textwidth]
{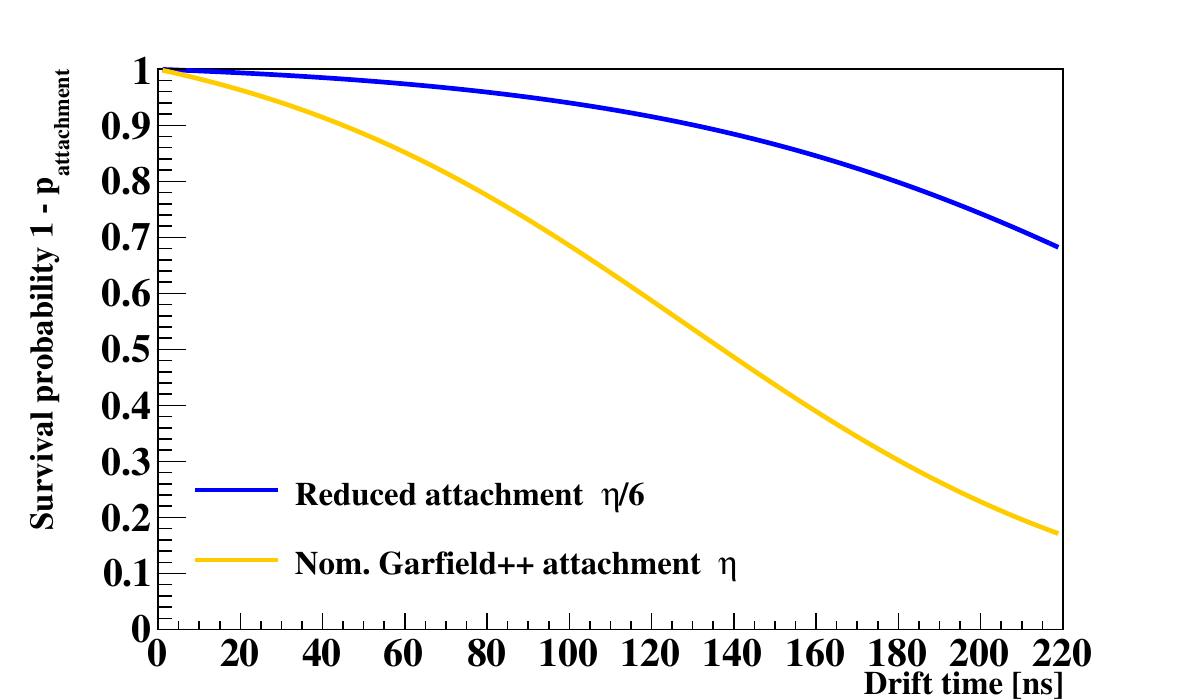}
\qquad
\includegraphics[width=.41\textwidth]
{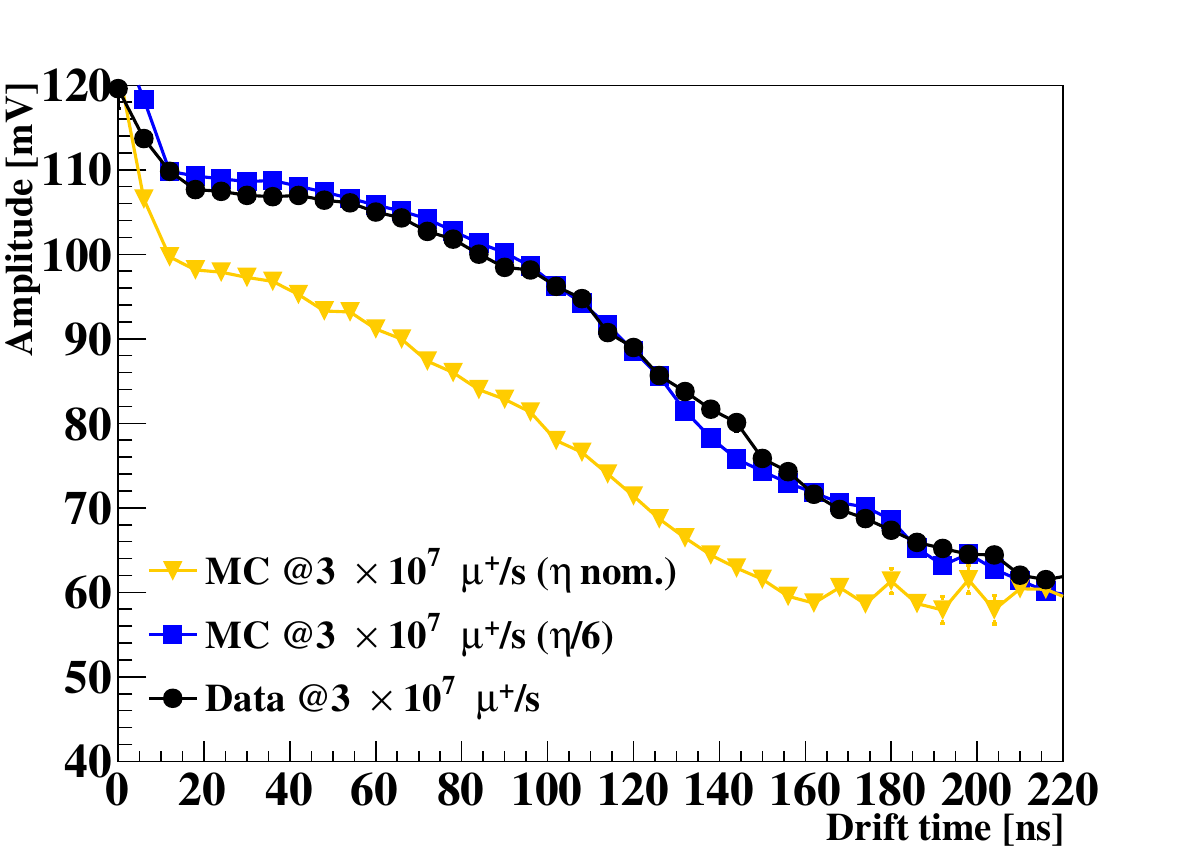}
\caption{Garfield++ predictions of cluster survival probability vs drift time (left) and comparison between experimental and computed cluster amplitude in the MEG II data (right). For the calculation the attachment coefficient was reduced by a factor six. \label{fig:attachment}}
\end{figure}

However, the literal use of Garfield++ \cite{Garfieldpp_web}, even in drift cells as small as those in MEG II, seems to predict a drift electron survival probability of around \SI{40}{\percent} for drift times corresponding to half the cell width (about 150 ns, or approximately 3.3 mm), as shown in Fig.~\ref{fig:attachment} on the left where this probability is derived from Garfield++ (orange curve).
In reality, the measurement of signal amplitude as a function of drift time (Fig.~\ref{fig:attachment} on the right) shows that the experimental data agree with an attachment value that is roughly reduced by a factor of six compared to that predicted by Garfield++. This results in an increase of the previously estimated electron survival probability at the edge of the cell to about \SI{90}{\percent} (Blue curve in Fig.~\ref{fig:attachment} left).

\section{A Dedicated Experiment to Measure the Attachment Coefficient in the MEG II Mixture}

\subsection{Time Projection Chamber}

Measuring a relatively large attachment coefficient $\eta$, with $1/\eta \sim O(\SI{1}{\centi\meter})$, requires a stable and well localized ionization inside a uniform electric field extending over a few centimeters. The number of electrons surviving at a given drift distance $d$ within the electric field will be given by:
\begin{equation}
    N(d) = N_0 e^{-\eta d}
\end{equation}
If a signal proportional to the number of electrons can be measured after the drift, the modulation of the signal as a function of the drift distance provides a measurement of the attachment coefficient.

\begin{figure}[htbp]
\centering
\includegraphics[width=\textwidth]{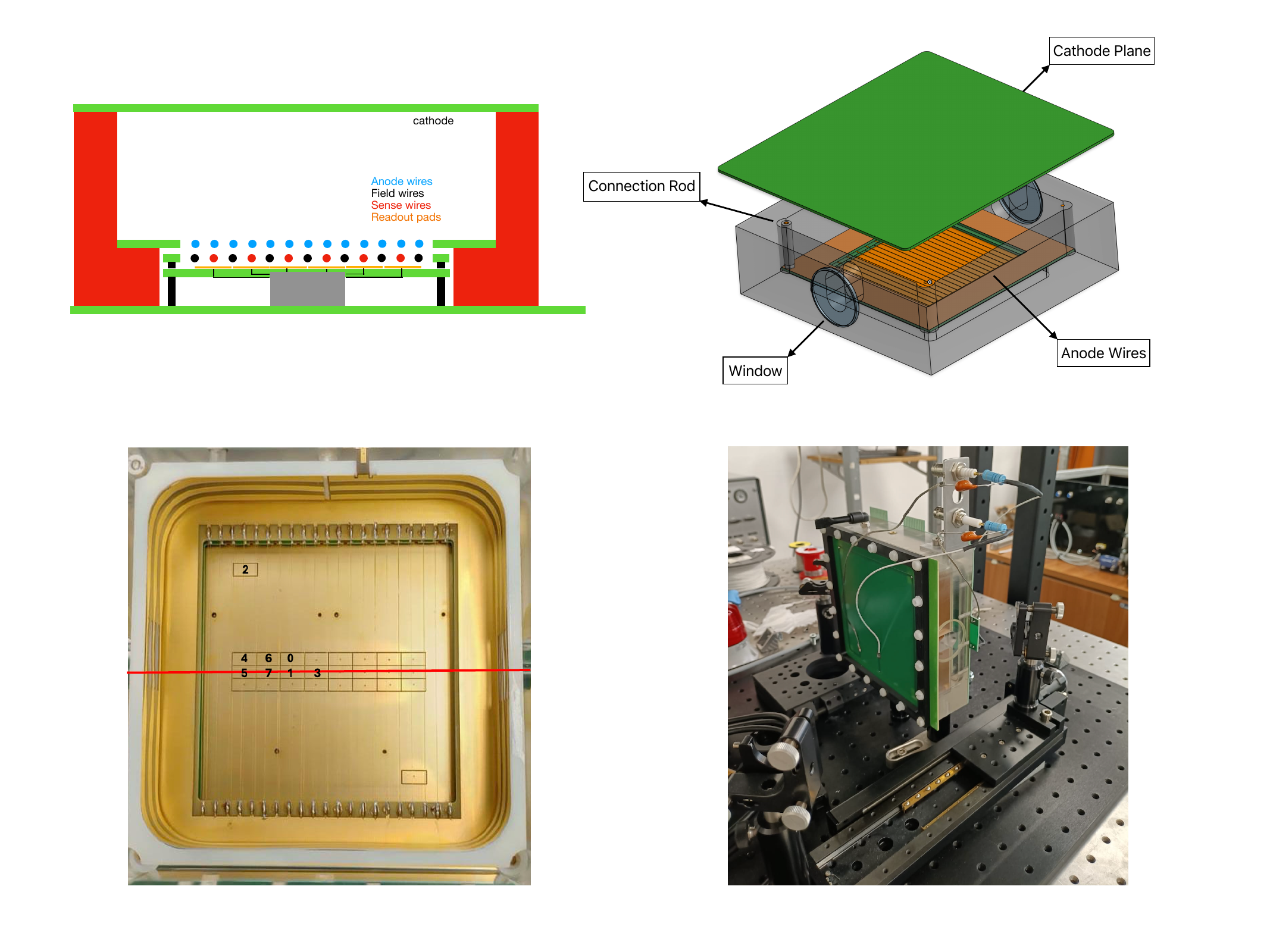}
\caption{\label{fig:tpc}A schematic of the TPC (top left, not in scale), a simplified drawing showing the cathode, the multiplication structure and the quartz windows (top right), a picture showing the readout pads below the wires, with our numbering scheme and the nominal trajectory of the laser beam (bottom left), and the TPC installed on the laser path (bottom right). Pad number 2 is the reference pad mentioned in Sec.~\ref{sec:data}.} 
\end{figure}

For this purpose, a small Time Projection Chamber (TPC) was built, with a \SI{3}{\centi\meter} drift region delimited on one side by a \SI{10}{}$\times$\SI{10}{\centi\meter\squared} cathode plane at high negative voltage, and on the other side by an arrangement of thin wires serving as the electron multiplication structure. One layer of \SI{80}{\micro\meter} wires at ground voltage separates the drift and multiplication regions (anode wires). A second layer of alternating \SI{80}{\micro\meter} field wires and \SI{25}{\micro\meter} sense wires, respectively at ground voltage and high positive voltage, creates the high field region which is necessary for the formation of the electron avalanche. This structure is closed by a plane at ground voltage, with its central part segmented into 24 rectangular, \SI{4}{}$\times$\SI{8}{\milli\meter\squared} pads which are connected to preamplifiers for the readout of the induced signals. The same frontend and DAQ electronics as in the MEG II drift chamber is used~\cite{Chiarello:2016nbj,Galli2019}, allowing to collect waveforms at \SI{0.7}{GSPS}. Due to the limited number of available readout channels, only eight pads could be read. The gas-tight body of the chamber is made in PMMA, with two quartz windows aligned to the center of the drift region. A sketch of the TPC with some pictures is shown in Fig.~\ref{fig:tpc}.

Ionization at constant intensity and at a given drift distance from the readout is produced by a pulsed \SI{355}{\nano\meter} laser beam via multi-photon ionization. The laser beam goes through the quartz windows, travels parallel to the cathode and wire planes, and is focused approximately at the center of the TPC. A set of mirrors and a micrometer stage allow to align the laser beam and to move it at different drift distances. Due to the small dimension of the readout pads, the signal induced by a single avalanche is shared among multiple contiguous pads, with sharing ratios depending on the position of the avalanche. For this reason, if the beam is not kept well centered with respect to the pads themselves once it is moved at different drift distances, a modulation of the signal in a single pad would be observed, which could mimic the effect of the attachment. In order to mitigate these effects, three alignment targets mounted on the TPC body are used to ensure the correct alignment of the beam, with an accuracy better than \SI{1}{\milli\meter}. Fig.~\ref{fig:tpc} also shows the TPC installed on the laser path.

\subsection{Data samples and analysis}
\label{sec:data}

We produced gas mixtures of Helium and Isobutane with variable concentrations of Oxygen and Isopropyl Alcohol using a set of mass flow controllers. The correctness of the Helium to Isobutane volume ratio (90:10) was verified within \SI{2}{\percent} relative uncertainty with a binary gas analyzer\footnote{Stanford Research Systems BGA244} with its factory calibration and an electronic gas density meter\footnote{Avenysense Northdome} calibrated against a certified mixture. The Oxygen concentration was measured using an Oxygen analyzer with a resolution of \SI{0.01}{\percent}, calibrated against ambient air. As in the MEG II experiment, Alcohol is added to the mixture by sending a known fraction of the gas flow through a bubbler, and the Alcohol concentration is estimated by measuring the Alcohol consumption with a precision scale.

Data were collected with drift fields from \SI{700}{\volt\per\centi\meter} to \SI{1500}{\volt\per\centi\meter}. The energy of the beam pulses was set around \SI{50}{\micro\joule} and the amplification voltage applied to the sense wires was between \SI{1220}{\volt} and \SI{1260}{\volt}. At higher beam intensities or amplification voltages, the large size of the avalanches was observed to generate saturation effects due to the screening potential of space charges, compromising the linearity of the detector response. 

Under these conditions, the amplified signals do not clearly emerge over the noise, which was typically around \SI{80} mV peak-to-peak in a \SI{1.5}{\micro\second} acquisition window. However, the noise is mostly coherent over all channels, and one of the readout pads (reference pad hereafter) was positioned far from the path of the laser beam, to be free from real signals and hence usable for noise subtraction purposes. For these reasons, the analysis of the signals proceeded as follows.
\begin{enumerate}
\item 1000 events were collected for each setting of drift distance and drift field, with the acquisition triggered by the same signals which triggers the laser beam pulse;
\item for each event, the waveform of the reference pad is subtracted from the waveform of any other channel, and a low-pass digital filter is applied with a cutoff at 150 \SI{}{\mega\hertz};
\item all noise-subtracted waveforms for a given channel are summed, in order to further suppress the random noise;
\item the average size $Q$ of the signals, proportional to the average number of electrons reaching the multiplication stage, is estimated as the integral of the summed waveform in a range of [-50,+250] digitization samples (which corresponds approximately to [-70,+360]~ns) around the signal's leading edge.
\end{enumerate}

\begin{figure}[htbp]
\centering
\includegraphics[width=\textwidth]{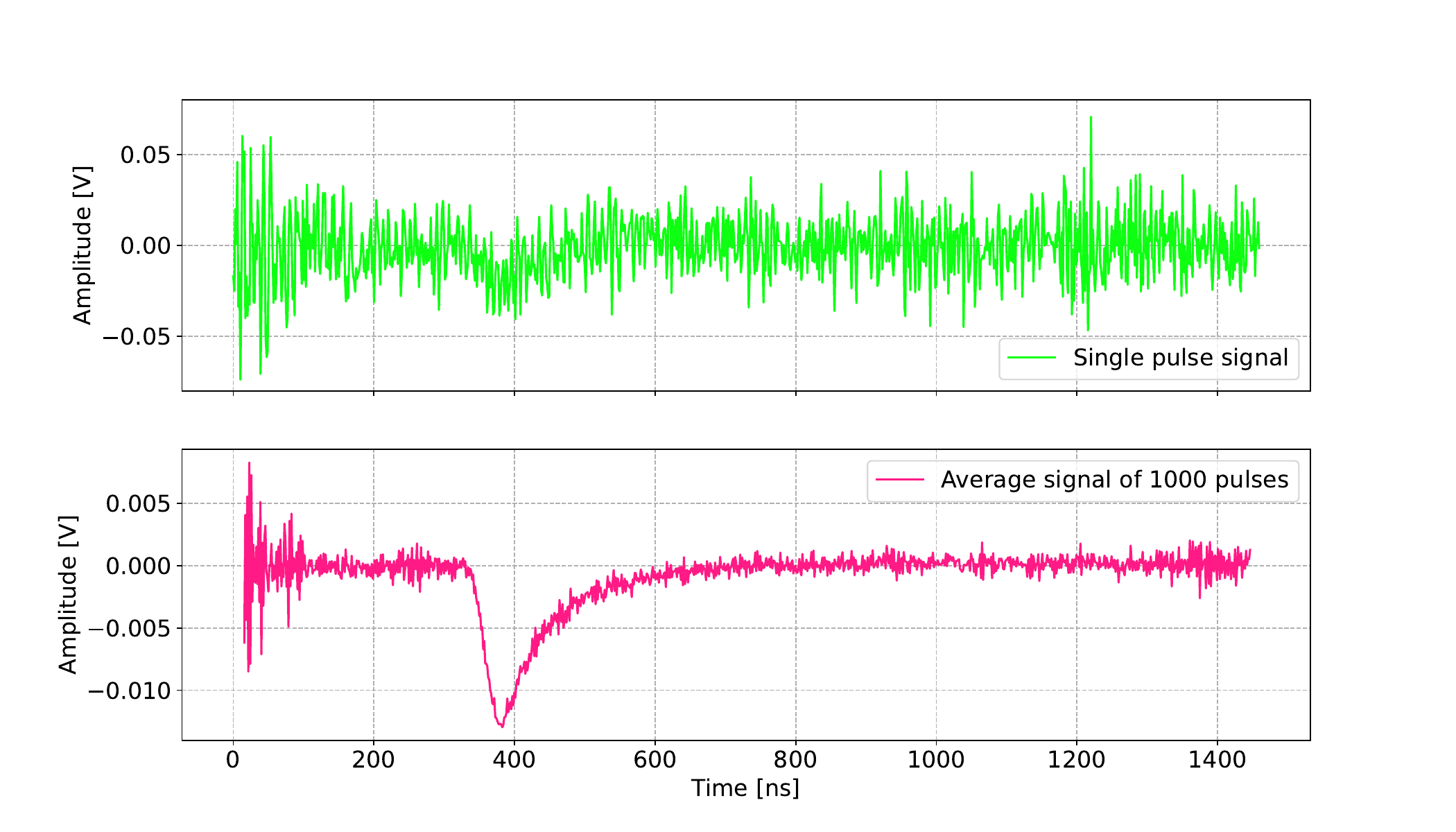}
\caption{\label{fig:signal} Top: an example of waveform from channel 3 (the signal is barely visible around 400 ns). Bottom: the average signal over 1000 events after subtracting the common noise and filtering out the high-frequency components.} 
\end{figure}

In Fig.~\ref{fig:signal} examples of a raw signal and an average signal over 1000 events are shown.

\subsection{Results}

\begin{figure}[htbp]
\centering
\includegraphics[width=0.8\textwidth]{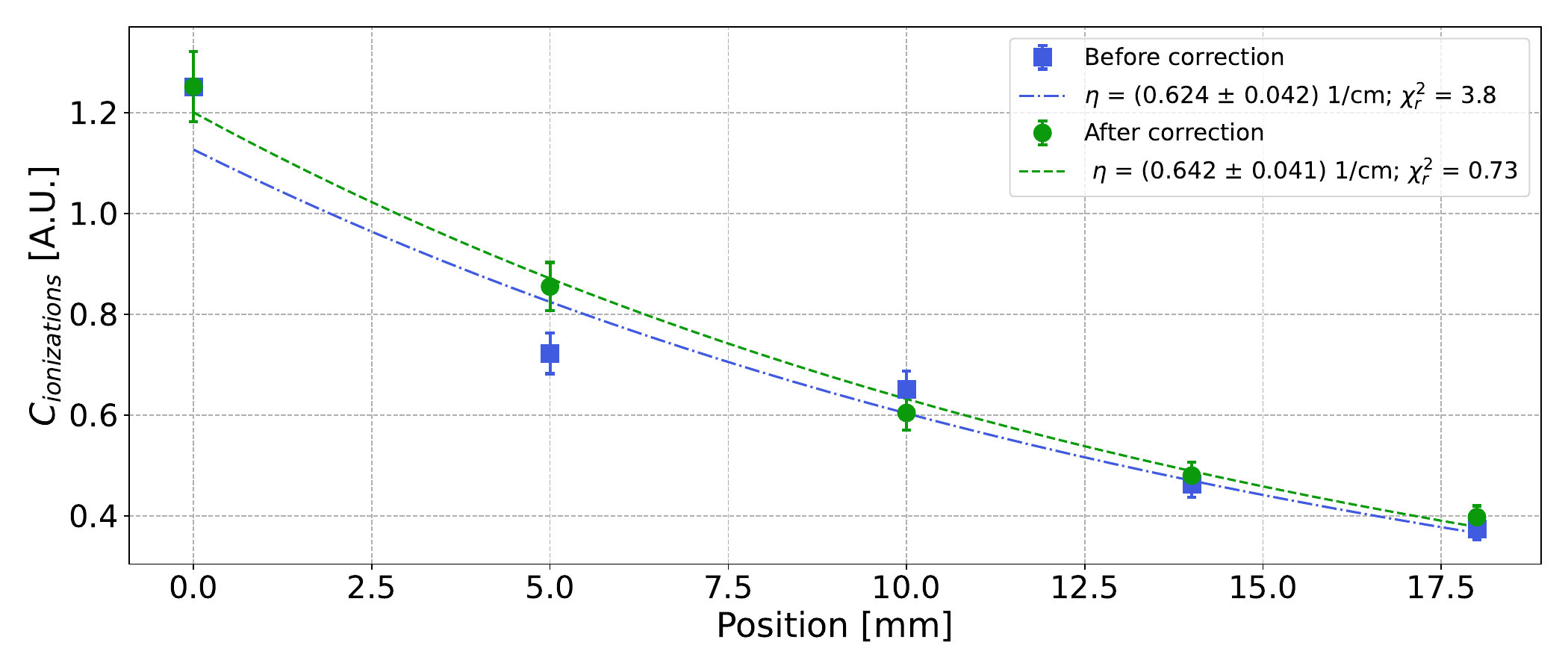}
\includegraphics[width=0.8\textwidth]{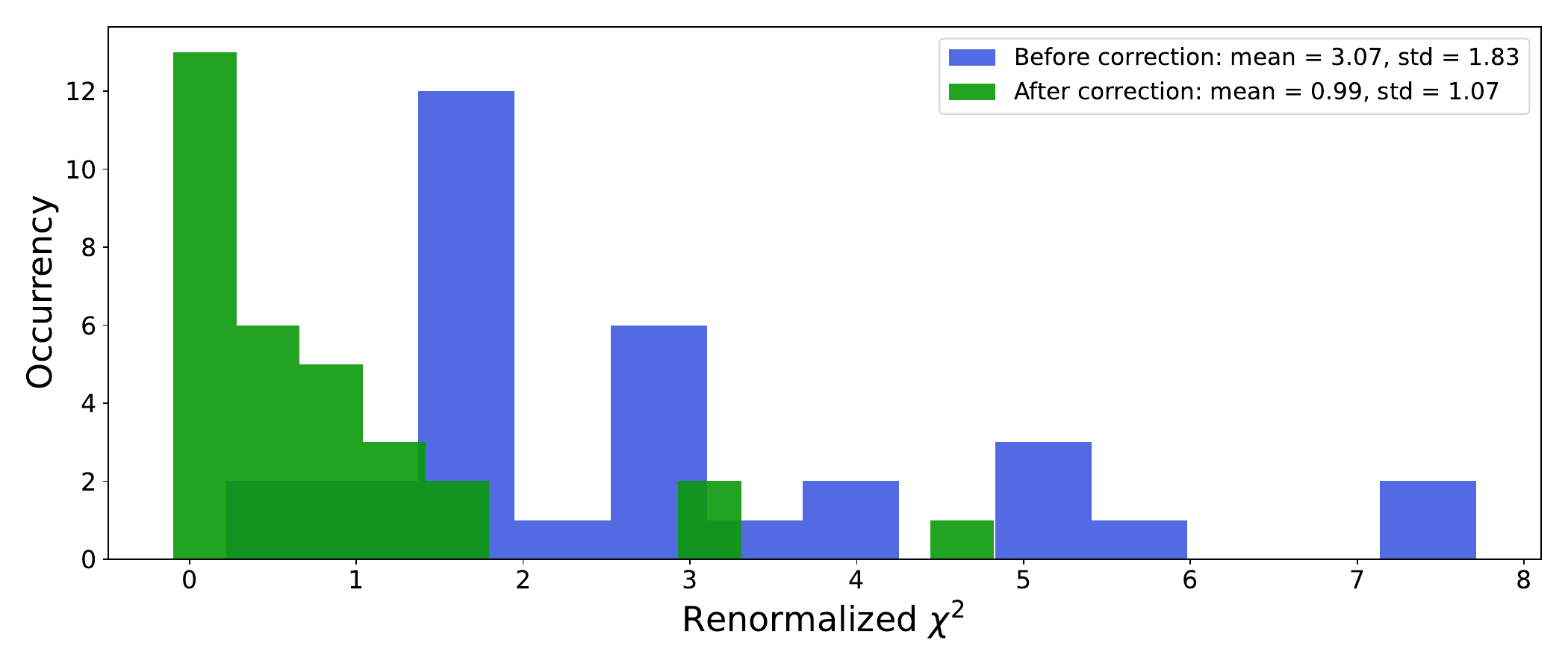}
\caption{\label{fig:attachment_correction} Top: signal size $Q$ as a function of the drift distance before and after the correction described in the text is applied. Bottom: distribution of the $\chi^2$ of the exponential fits before and after the correction is applied (errors are chosen to get a unit average of the distribution after the correction).}
\end{figure}

In presence of Oxygen in the mixture, we observe that the signal size varies with a reasonably exponential trend as a function of the drift distance. However, comparing trends observed under different conditions, some systematic deviations were found, which could be ascribed to some  disuniformities in the transmitted laser beam intensity at different drift distances (for instance due to some defects in windows and mirrors). For this reason, data collected without Oxygen and Alcohol, which are known to have negligible levels of attachment according to~\cite{Golavatyuk2001} and hence the same signal size at any distance, are used to correct all the other measurements. In particular, given the signal sizes $Q_0(d_0)$ and $Q_0(d)$ measured without additives, respectively at the shortest drift distance $d_0$ and at distance $d$, the signal size $Q(d)$ measured in any other mixture is corrected as $Q^\prime(d) = Q(d) \cdot Q_0(d_0)/Q_0(d)$. Fig.~\ref{fig:attachment_correction} shows how the correction work in a specific case, and the improvement in the distribution of the $\chi^2$ over all the fits after the correction is applied.

In order to account for unknown sources of uncertainties, the error on the signal size $Q$ is determined in such a way that the distribution of the $\chi^2$ over all fits, after the correction above is applied, has approximately unit average.

\begin{figure}[htbp]
\centering
\includegraphics[width=0.49\textwidth]
{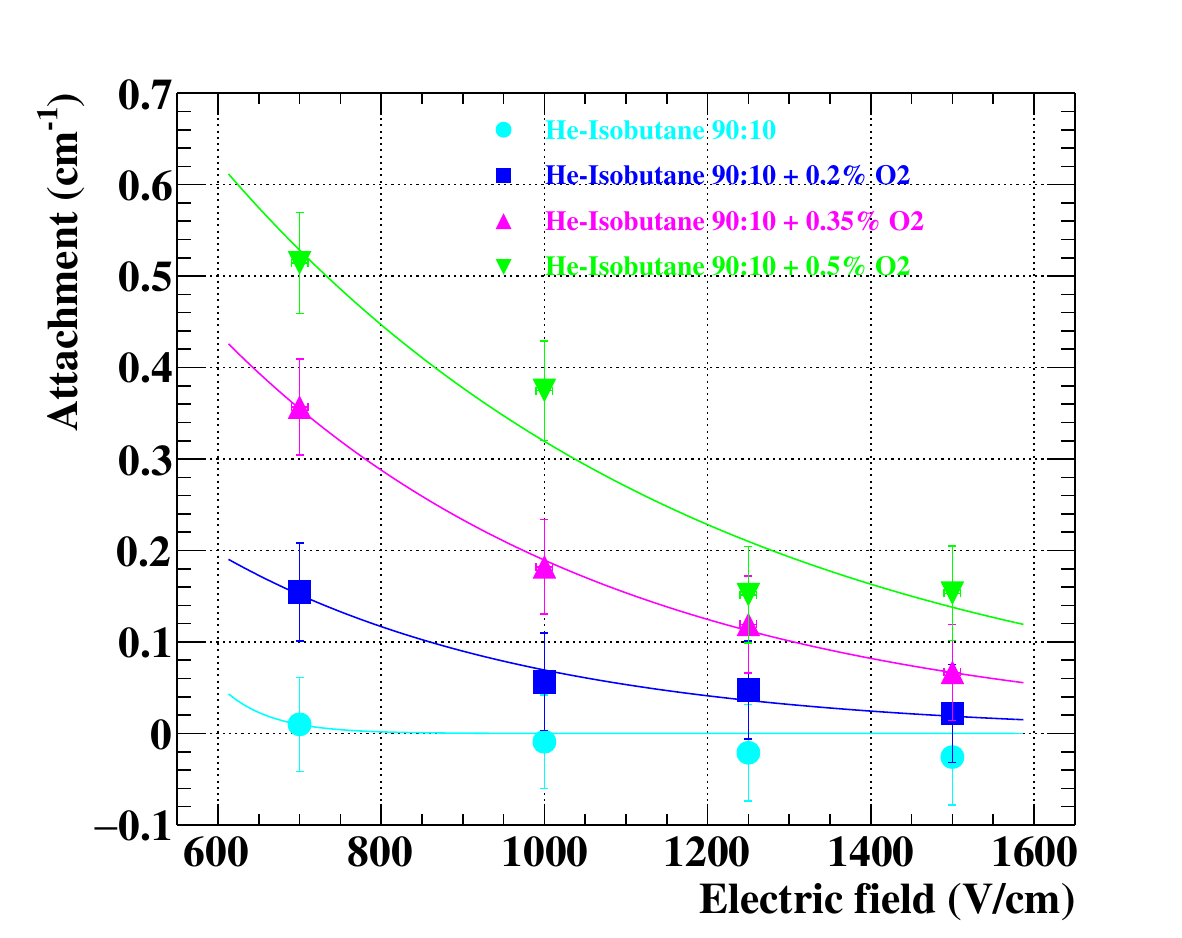}
\includegraphics[width=0.49\textwidth]
{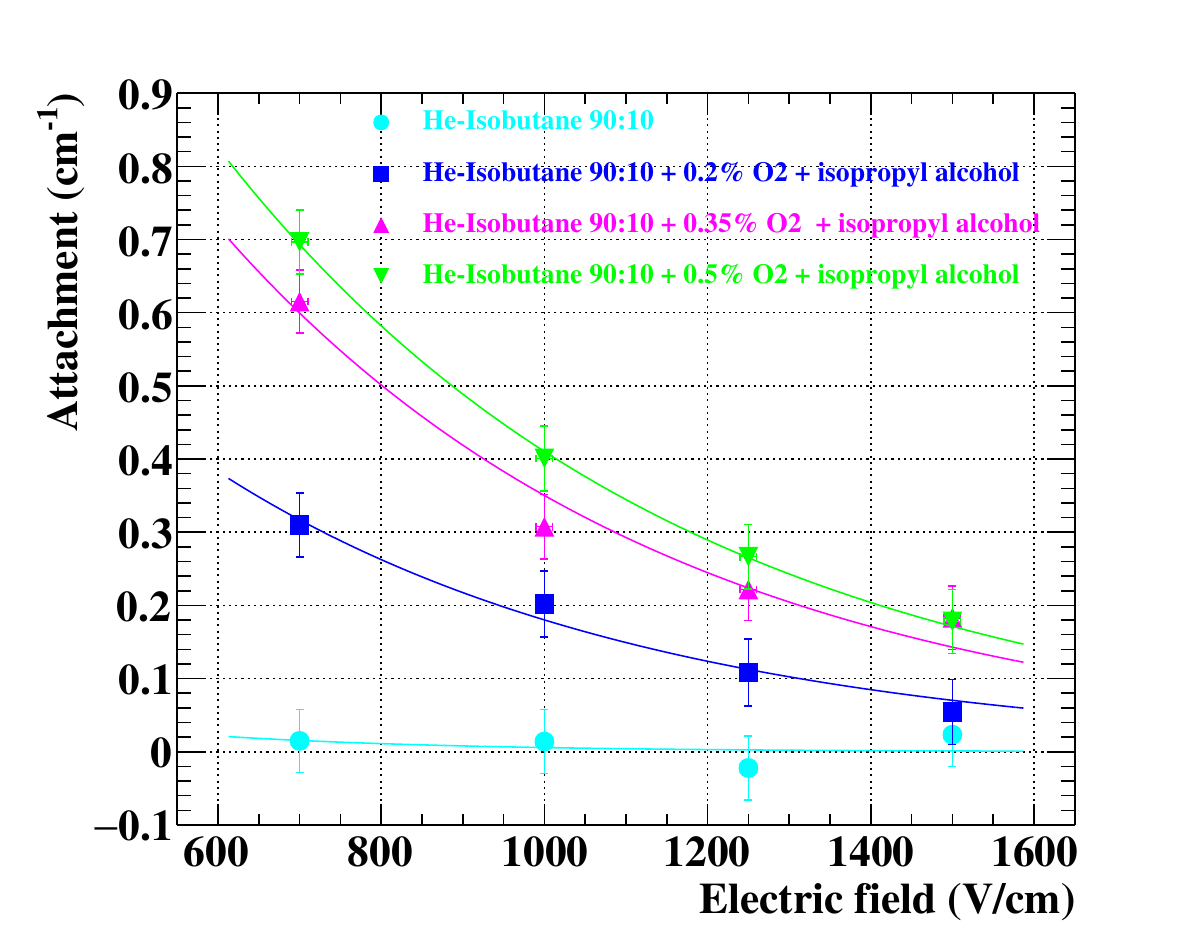}
\caption{\label{fig:attachment_all}Attachment coefficient $\eta$ measured at various concentrations of additives as a function of the drift field, with superimposed exponential fits for eye guidance. Mixture compositions are reported in the plots.} 
\end{figure}

Fig.~\ref{fig:attachment_all} shows the attachment coefficient $\eta$ measured at various concentrations of additives as a function of the drift field. The plot on the left shows the attachment coefficient measurements for mixtures with various Oxygen concentrations and no Alcohol, that on the right the corresponding measurements for the same Oxygen contents but including Alcohol. Possible systematic uncertainties related to residual saturation effects were assessed by comparing results obtained with different multiplication voltages. Given the original voltage $U$, all measurements were repeated with \SI{20}{\volt} larger voltage, and we studied the distribution of the differences in $\eta$ divided by their uncertainties summed in quadrature:
\begin{equation}
\frac{\eta(U) - \eta(U+\SI{20}{V})}{\sigma_\eta(U) \oplus \sigma_\eta(U + \SI{20}{V})}
\end{equation}
In a non-saturated regime, $\eta$ would be independent of the multiplication voltage, and the distribution would be centered at zero. The measured distribution was observed to be centered at 0.2 (i.e. the difference is, on average, \SI{20}{\percent} of the uncertainty). It could indicate that $U+\SI{20}{V}$ is not low enough to be in a non-saturated regime. Although $U$ could be low enough, as our preparatory linearity studies were suggesting, we increased all uncertainties, to be conservative, by \SI{20}{\percent}.

\begin{figure}[htbp]
\centering
\includegraphics[width=0.49\textwidth]
{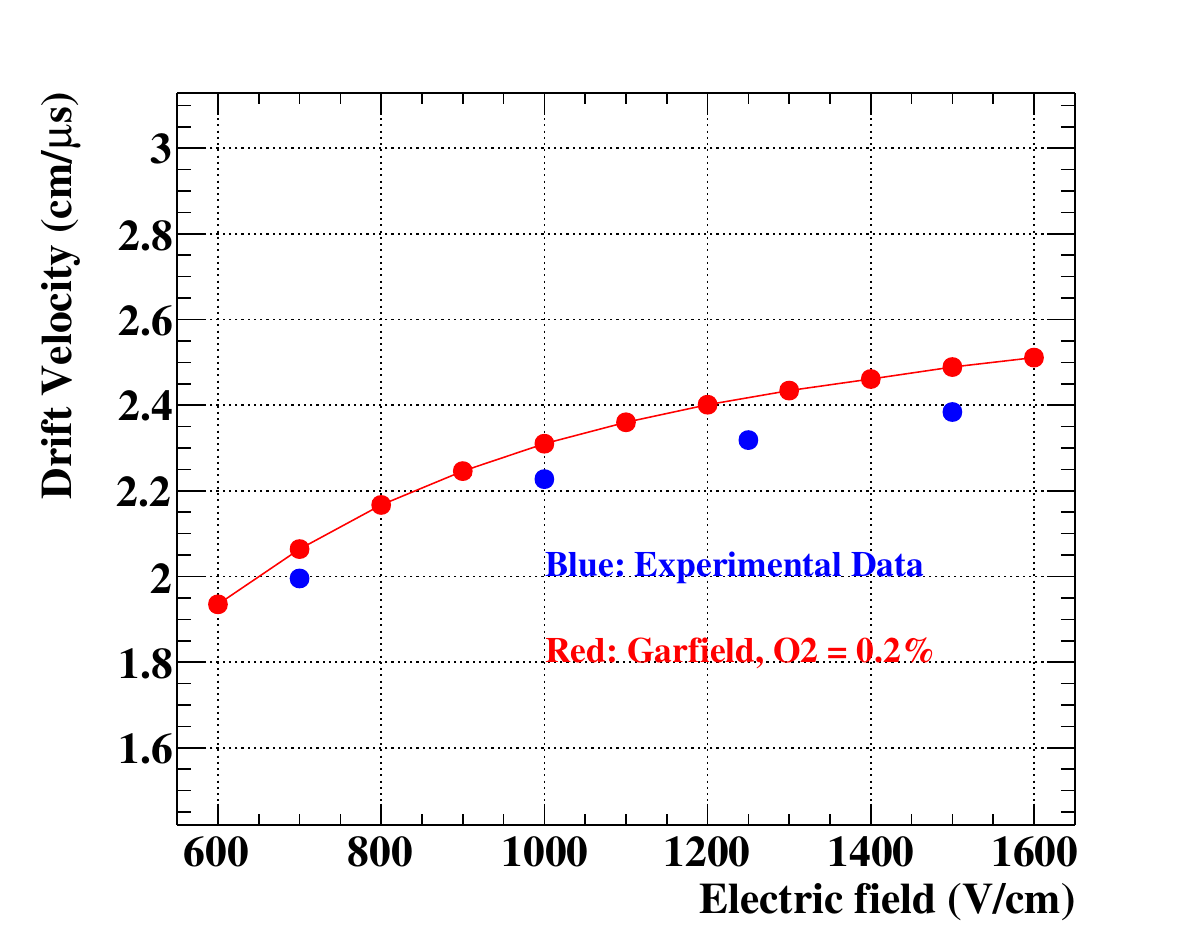}
\includegraphics[width=0.49\textwidth]
{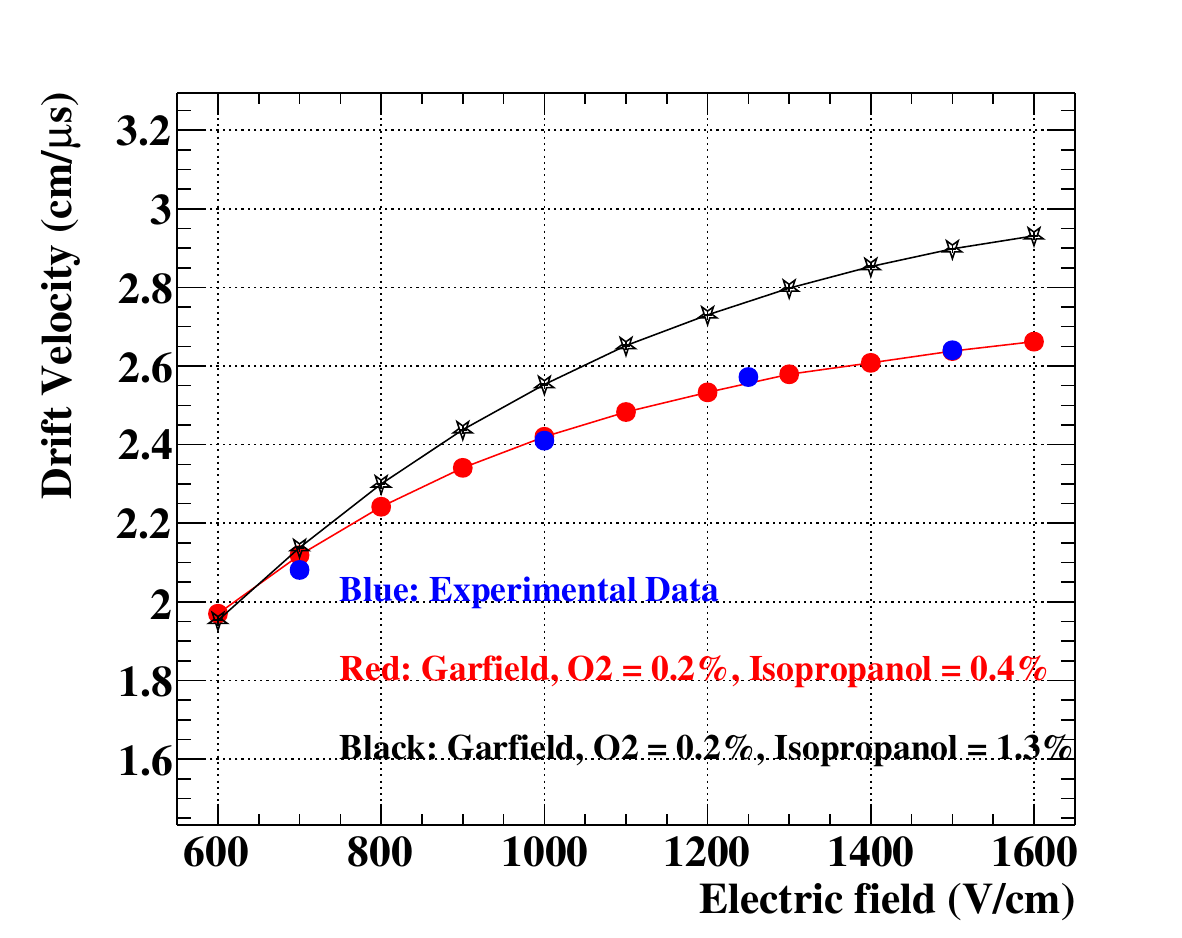}
\caption{Comparison between the measured and the Garfield++ computed drift velocity as a function of the electric field for the MEG II gas mixture Helium:Isobutane (90:10) with only \SI{0.2}{\percent} Oxygen (left) or with \SI{0.2}{\percent} Oxygen and Alcohol (right). The results for different Oxygen concentrations do not vary since Oxygen content does not affect significantly the drift velocity. Since the Alcohol concentration is only measured indirectly by means of the consumption rate, the Alcohol concentration is known with a large uncertainty and two representative Alcohol concentrations (\SI{0.4}{\percent} and \SI{1.3}{\percent}) are used in the Garfield++ simulation. The better agreement is obtained by using \SI{0.4}{\percent}, while the experimental data are a few per cent lower when one uses the nominal Alcohol concentration of \SI{1.3}{\percent}. \label{fig:velocity}}
\end{figure}

As a byproduct, we could also measure the electron drift velocities under the same conditions, which are a useful cross check of the data quality when compared with the simulations by Garfield++. The results are shown in Fig.~\ref{fig:velocity}. 

The measured values in the mixture without alcohol are lower than the Garfield++ predictions by about \SI{3}{\percent}. A similar deviation can be observed in~\cite{Cavoto:2023uvf}, where a completely different setup was used. However, in our present case, this systematic difference could be also generated by residual misalignments at the level of a few hundred micrometers. 

\section{Predictions from Garfield++ and Comparison with MEG II Measurements and Previous Measurements}

\subsection{Garfield++ Simulation of Attachment on Oxygen}

The electron attachment in the presence of Oxygen is described in Garfield++ by a two-stage mechanism:

\begin{equation}
    e^- + O_2 \rightarrow (O_2^-)^* \quad \text{(excited ion formation)}
\end{equation}

followed by de-excitation of the Oxygen ion through collision with another Oxygen molecule or another molecule X in the mixture:

\begin{equation}
    (O_2^-)^* + O_2 \rightarrow O_2^- + O_2 + \gamma \quad \text{(single process)}
\end{equation}
\begin{equation}
    (O_2^-)^* + X \rightarrow O_2^- + X + \gamma \quad \text{(three-body process)}
\end{equation}

This latter mechanism is called three-body because it involves the electron, Oxygen, and the molecule X.

The attachment rate \( R \) in a given mixture is proportional to the capture cross-section of Oxygen \( \sigma_C \), multiplied by the Oxygen number density in the mixture \( n_{O_2} \) (number of Oxygen nuclei per unit volume) and the de-excitation rate. The de-excitation rate is obtained by summing the de-excitation cross-sections \( \sigma_{d,X_i} \) for each of the components \( X_i \) in the mixture:

\begin{equation}
    R \propto \sigma_C \times n_{O_2} \times \sum_{i=1}^{N} \sigma_{d,X_i} \times n_{X_i}
\end{equation}

The density of element \( X_i \) can also be written as the density \( n \) of the gas mixture multiplied by the mass fraction \( f_{X_i} \) of component \( X_i \), and the de-excitation cross-section for each molecule \( X_i \) can be normalized to the one for Oxygen:

\begin{equation}
    R \propto \sigma_C \times n_{O_2} \times n \times \sigma_{d,O_2} \times \sum_{i=1}^{N} \frac{\sigma_{d,X_i}}{\sigma_{d,O_2}} \times f_{X_i}
\end{equation}

The behavior of each gas usable in Garfield++ is encoded in individual subroutines in the FORTRAN program magboltz.f. The capture and de-excitation rates for Oxygen are correctly calculated using the specified Oxygen density for the mixture, while the three-body process rate is calculated internally in the subroutine describing Oxygen, assuming the mixture is entirely composed of Oxygen. In fact, the parameter \( T3B \), defined as:

\[
T3B = \sum_{i=1}^{N} \frac{\sigma_{d,X_i}}{\sigma_{d,O_2}} \times f_{X_i}
\]

is normally set to one and must be modified by hand depending on the type of mixture used to obtain meaningful results. The most simple way to proceed is to modify the subroutine GAS15 in magboltz.f, which contains all cross sections, data statements etc. related to Oxygen. In this subroutine the default treatment of \( T3B \) parameter is fixed to 1, as observed:
\begin{verbatim}
C  THREE BODY ATTACHMENT
C *****************************************
C ENTER HERE SCALING FACTOR FOR THREE BODY ATTACHMENT IN MIXTURES:
C  FOR NORMAL SCALING T3B=1.0
C     T3B=1.0
C --------------------------------------------
\end{verbatim}.
This statement must be replaced by a different one, for instance as this, corresponding to the MEGII mixture: 
\begin{verbatim}
C  THREE BODY ATTACHMENT
C *****************************************
     T3B = RHE*FHE + RISO*FISO + RALC*FALC + ROXY*FOXY
C --------------------------------------------
\end{verbatim}
where FHE, FISO, FALC and FOXY are the molar fractions of Helium, Isobutane, Isopropanol and Oxygen of the mixture and RHE, RISO, RALC and ROXY are the ratios of the de-excitation cross sections for Helium, Isobutane, Isopropanol and Oxygen with respect to that of Oxygen (so, ROXY$ = 1$). The ratios RHE, RISO, RALC and ROXY can be computed once at the beginning of code execution using the values reported in literature.   
However, de-excitation cross-sections as a function of energy are known only for a limited set of gases \cite{Shimamori1977}. In the case of MEG II mixture, while there are measurements for Helium and Isobutane, those for Isopropanol are absent 
and the contribution of Isopropanol can be roughly estimated by using the mean of Ethanol and Methanol, whose experimental data are known\footnote{Stephen Biagi, private suggestion}. 

\subsection{Comparison of Garfield++ Attachment Estimates with Experimental Measurements}

\begin{figure}[htbp]
\centering
\includegraphics[width=0.48\textwidth]
{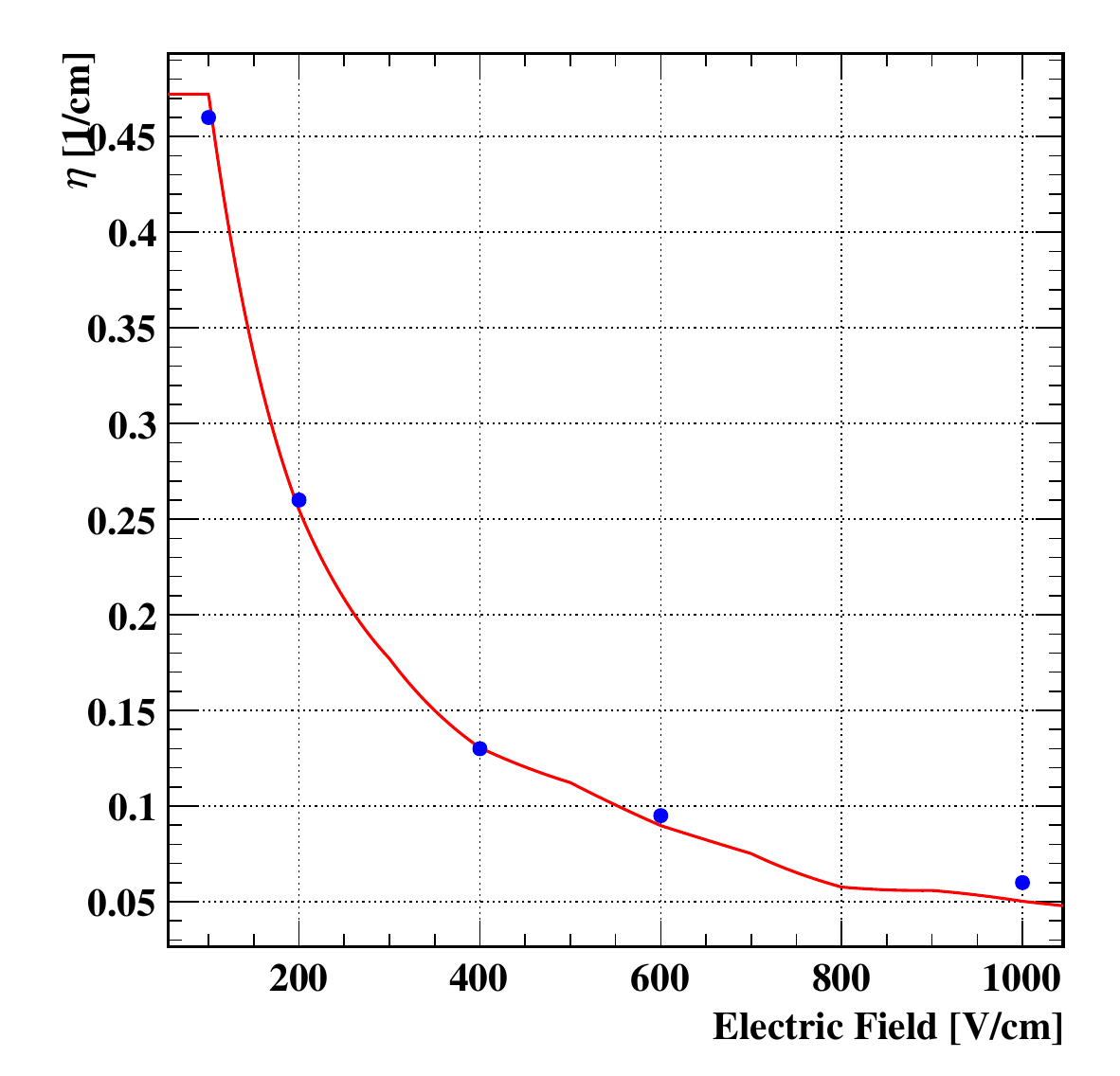}
\includegraphics[width=0.48\textwidth]
{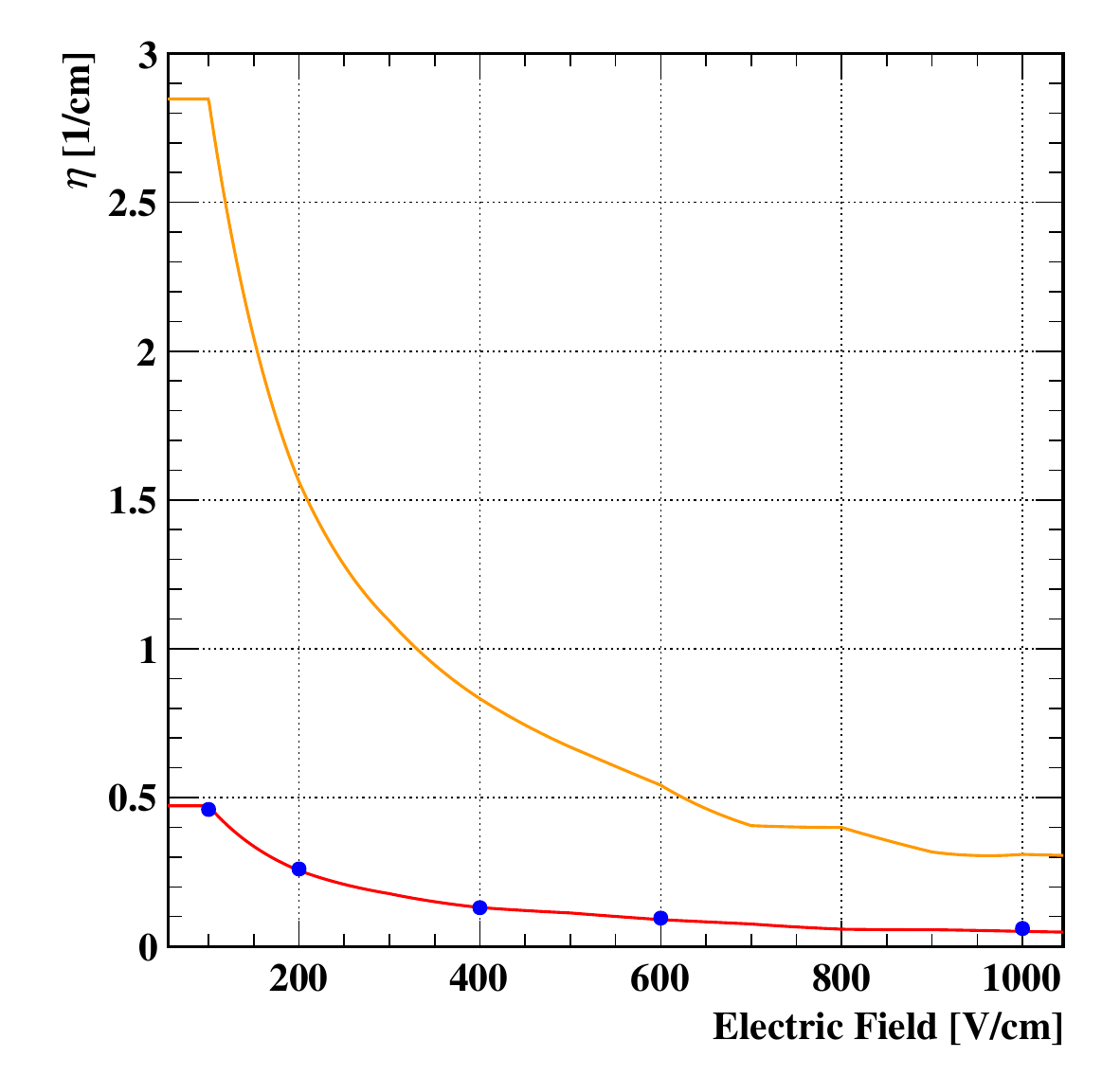}
\caption{Comparison between the attachment 
coefficient measured in paper \cite{Golavatyuk2001} and the Garfield++ predictions. The gas mixture is Helium:Isobutane 90:10 with 600 ppm of Oxygen; the Isobutane contribution to attachment cross section is reduced by \SI{20}{\percent} with respect to the value reported in \cite{Shimamori1977}, as explained in the text. The plot on the left shows the comparison between experimental points (blue points) and modified Garfield++ calculation (red curve), that on the right shows the difference between the modified (red curve) and the unmodified Garfield++ calculation (yellow curve).   \label{fig:Golavatyukcomp}}
\end{figure}

\begin{figure}[htbp]
\centering
\includegraphics[width=0.49\textwidth]{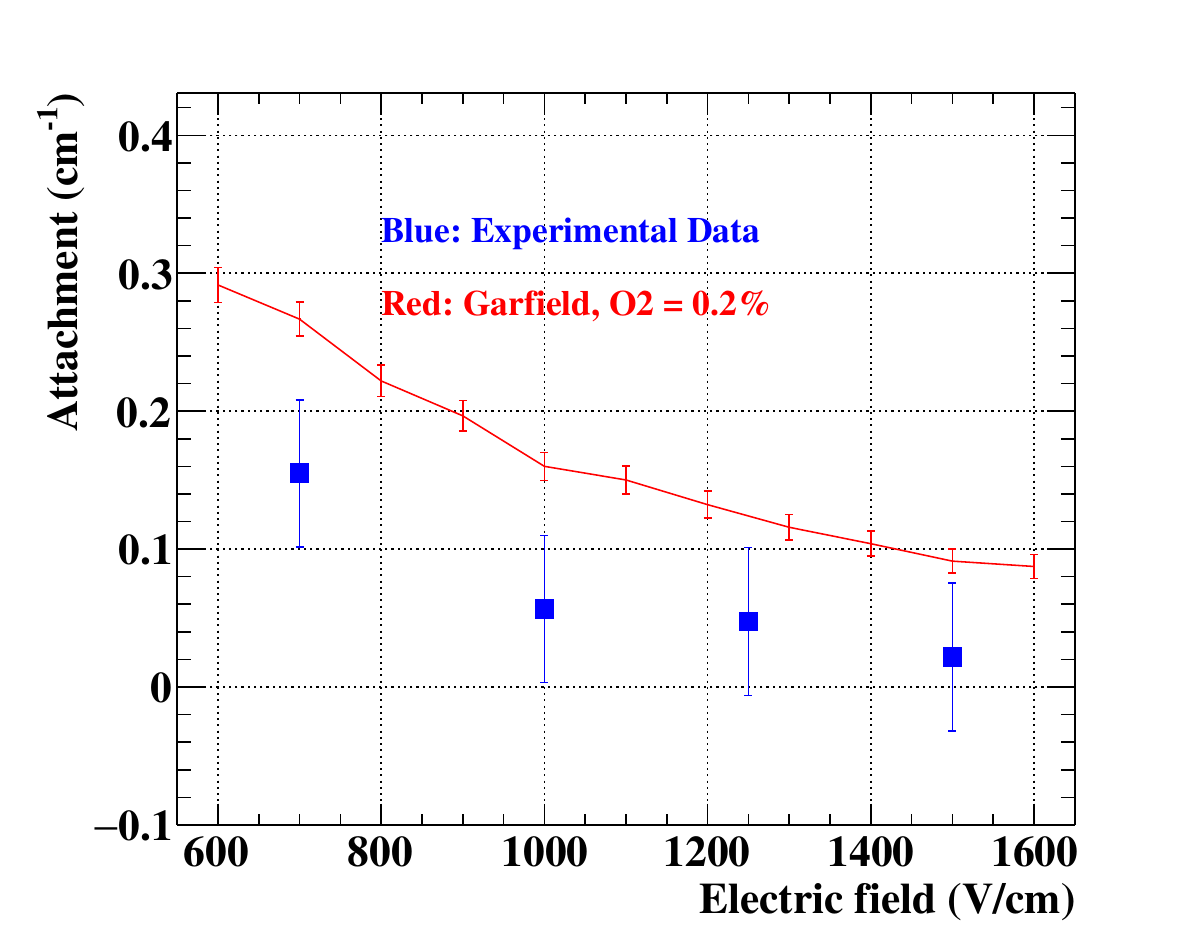}
\includegraphics[width=0.49\textwidth]{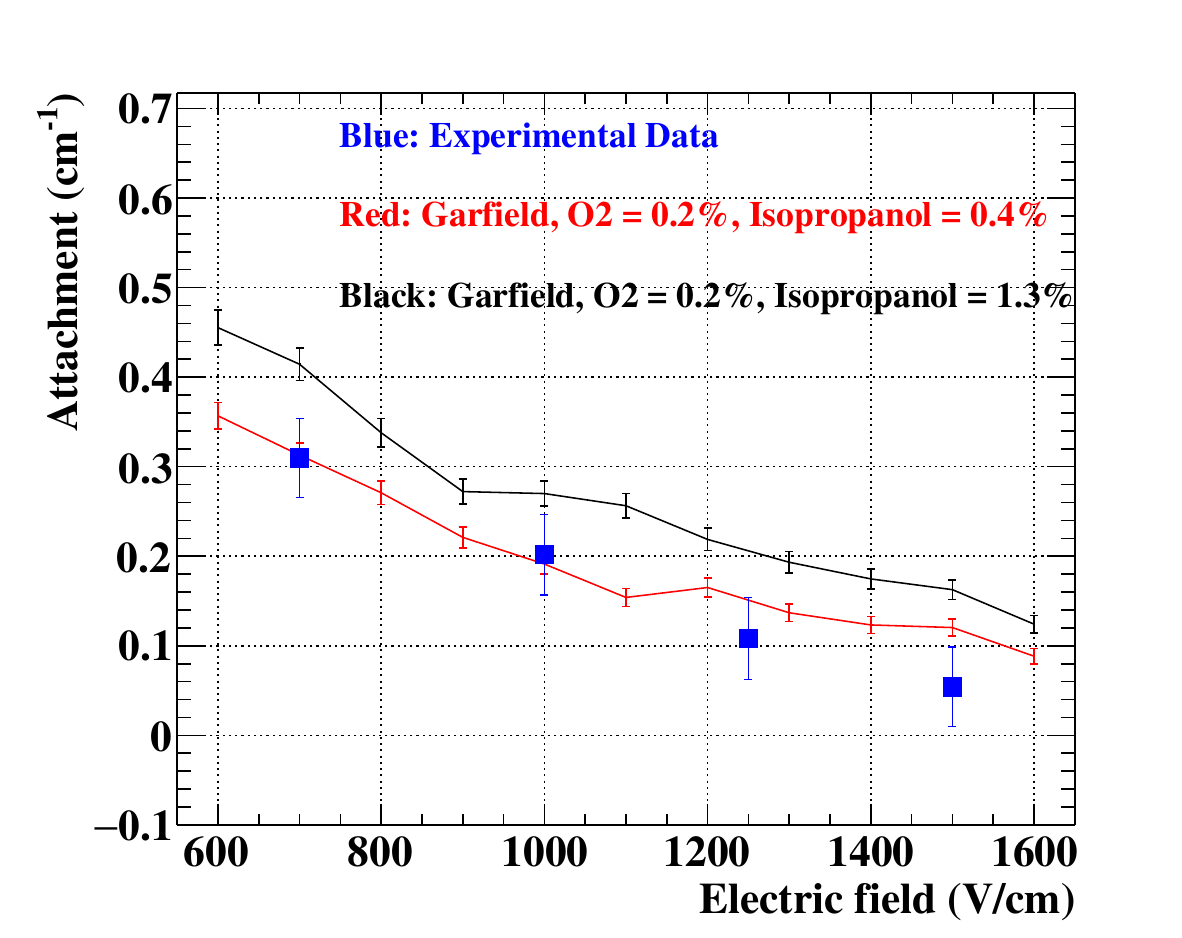}
\includegraphics[width=0.49\textwidth]{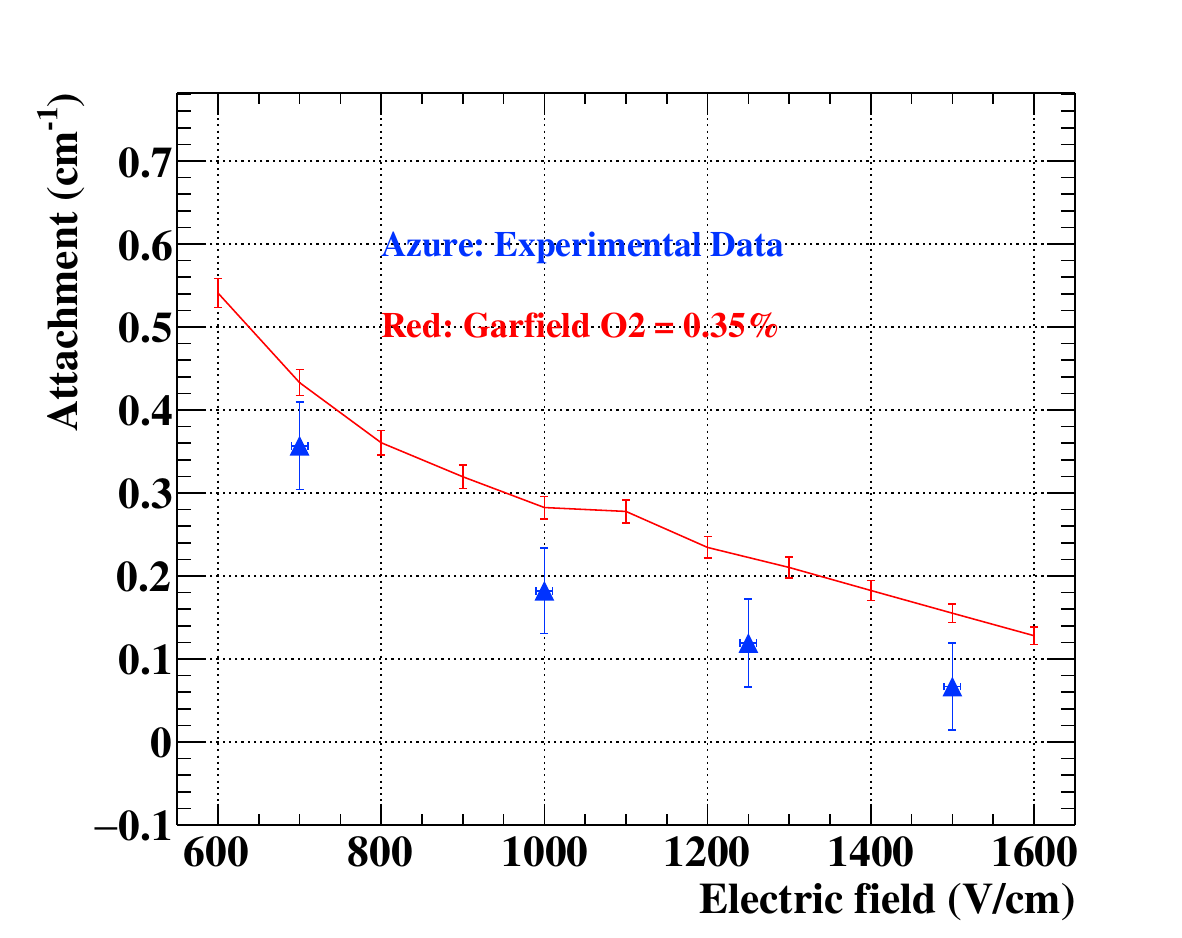}
\includegraphics[width=0.49\textwidth]{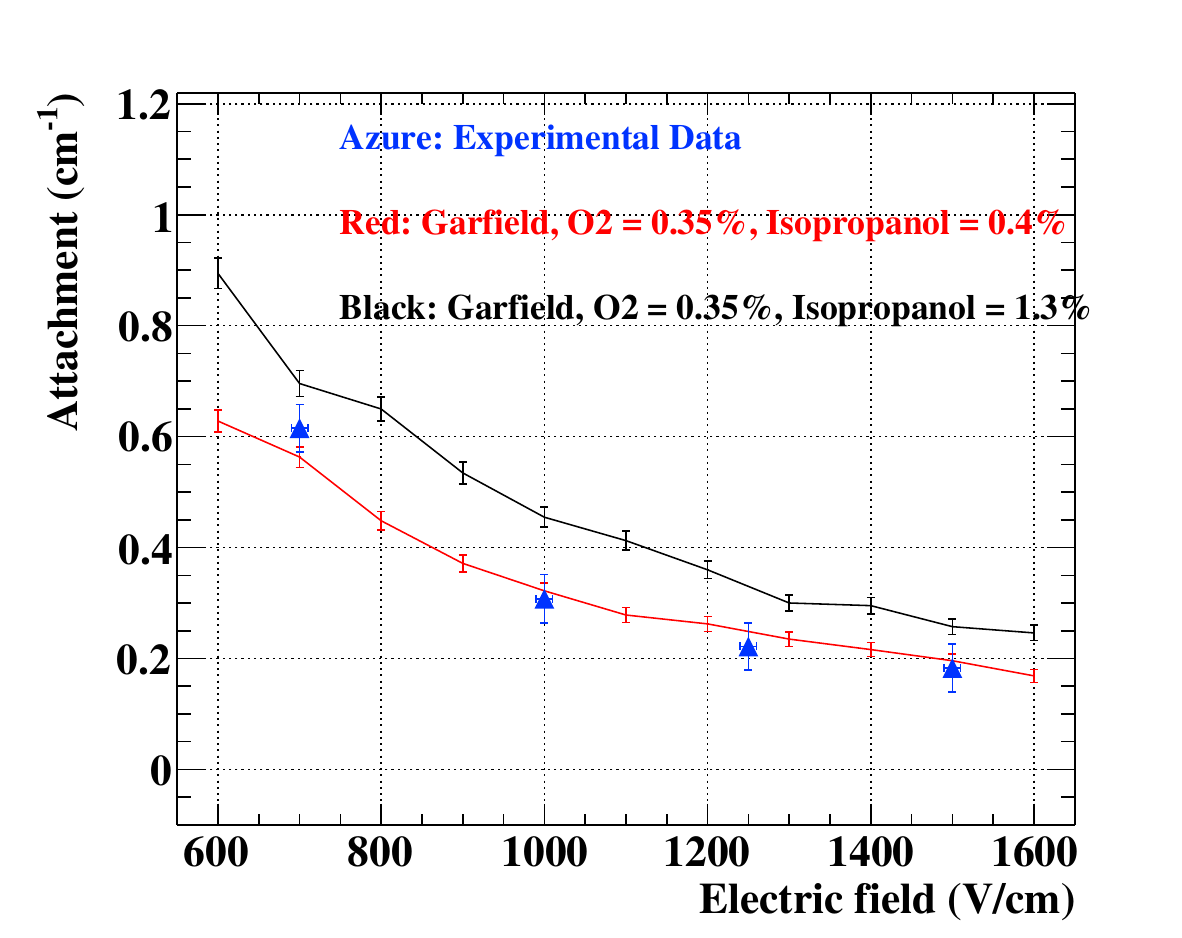}
\includegraphics[width=0.49\textwidth]{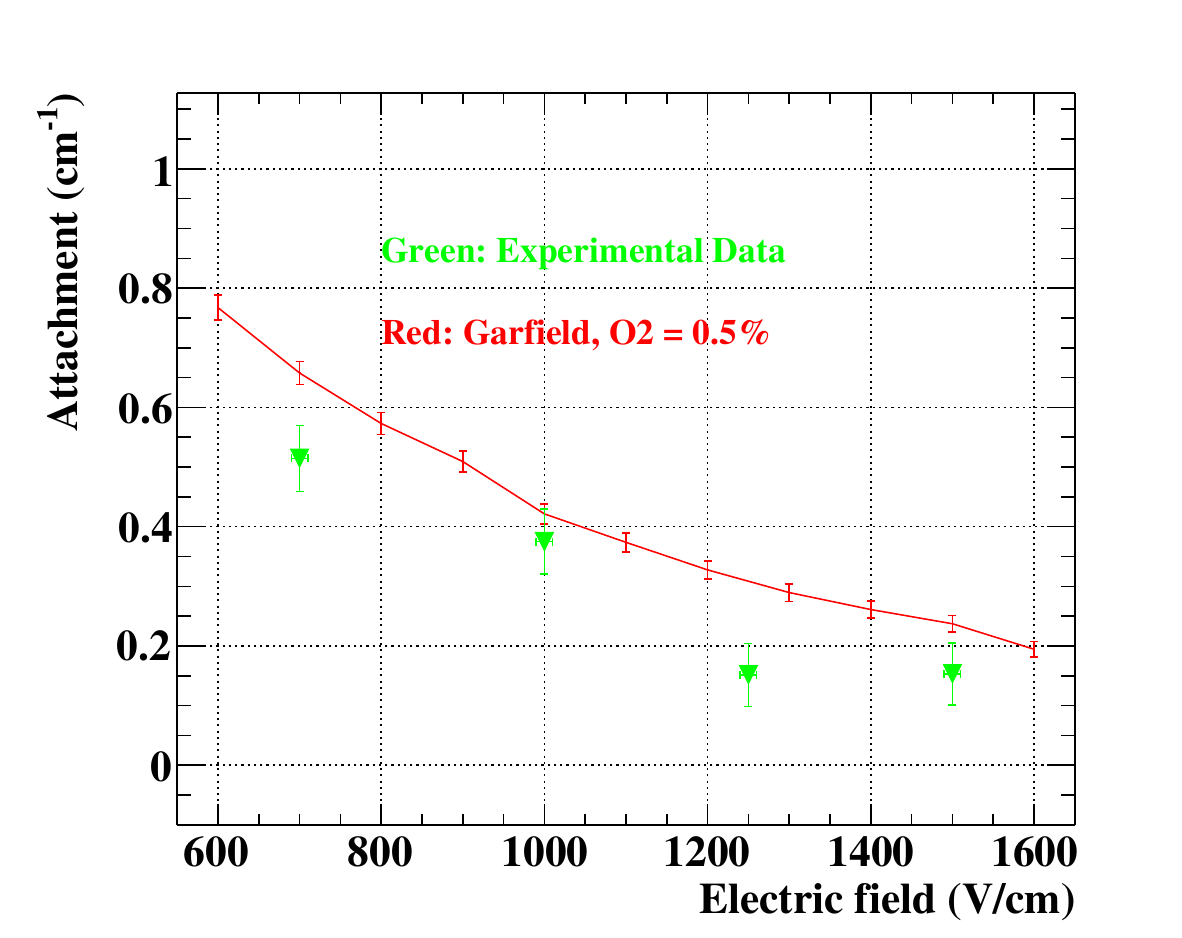}
\includegraphics[width=0.49\textwidth]{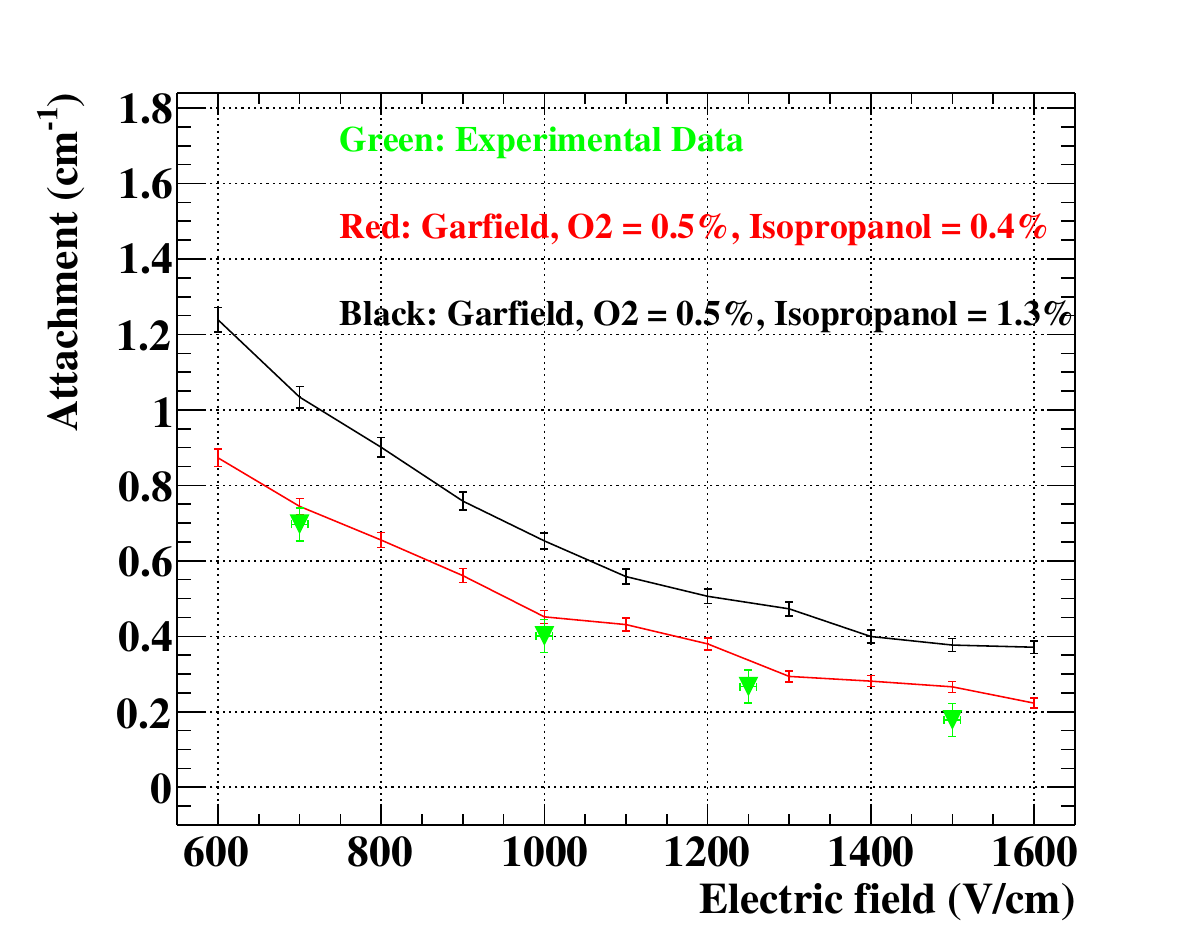}
\caption{Comparison between the measured and calculated electron attachment coefficient as a function of the electric field for three Oxygen concentrations: from top to bottom \SI{0.2}{\percent}, \SI{0.35}{\percent}, \SI{0.5}{\percent}. Plots on the left correspond to mixtures without Isopropyl Alcohol, that on the right to mixtures with Isopropyl Alcohol. In the Garfield++ calculation two representative values of Alcohol concentration were used, \SI{0.4}{\percent} and \SI{1.3}{\percent}. As for drift velocity, the \SI{0.4}{\percent} Alcohol concentration seems the best choice to reproduce the experimental measurements. Symbols and colors used to distinguish the various series of measurements are the same in Fig.~\ref{fig:attachment_all}, except for \SI{0.35}{\percent} Oxygen concentration where Magenta was replaced by Azure to avoid confusion with the Red line.  \label{fig:AllRomeComp}}
\end{figure}

Garfield++ estimates, using the appropriate \( T3B \) values, of the attachment coefficient as a function of electric field are compared in Fig.~\ref{fig:Golavatyukcomp} with experimental measurements obtained previously \cite{Golavatyuk2001} for a Helium:Isobutane 90:10 mixture with 600 ppm of Oxygen, and in Fig.~\ref{fig:AllRomeComp} for the MEG II mixture, which contains Isopropyl Alcohol, at various Oxygen percentages obtained through the setup described in Section 2.

The paper \cite{Shimamori1977} provides only a rough estimate of de-excitation cross-section for Isobutane, without any precise uncertainty statement. To achieve the good agreement presented the value of Isobutane de-excitation cross-section was empirically tuned and reduced by \SI{20}{\percent} with respect to \cite{Shimamori1977} paper. The accuracy of this choice was confirmed by comparing attachment measurements with Garfield tuned predictions for different gas mixtures containing Isobutane, as discussed in the next section.

\subsection{Comparison of Simulation with Existing Literature Data for Isobutane and Oxygen Mixtures}

\begin{figure}[htbp]
\centering
\includegraphics[width=0.6\textwidth]{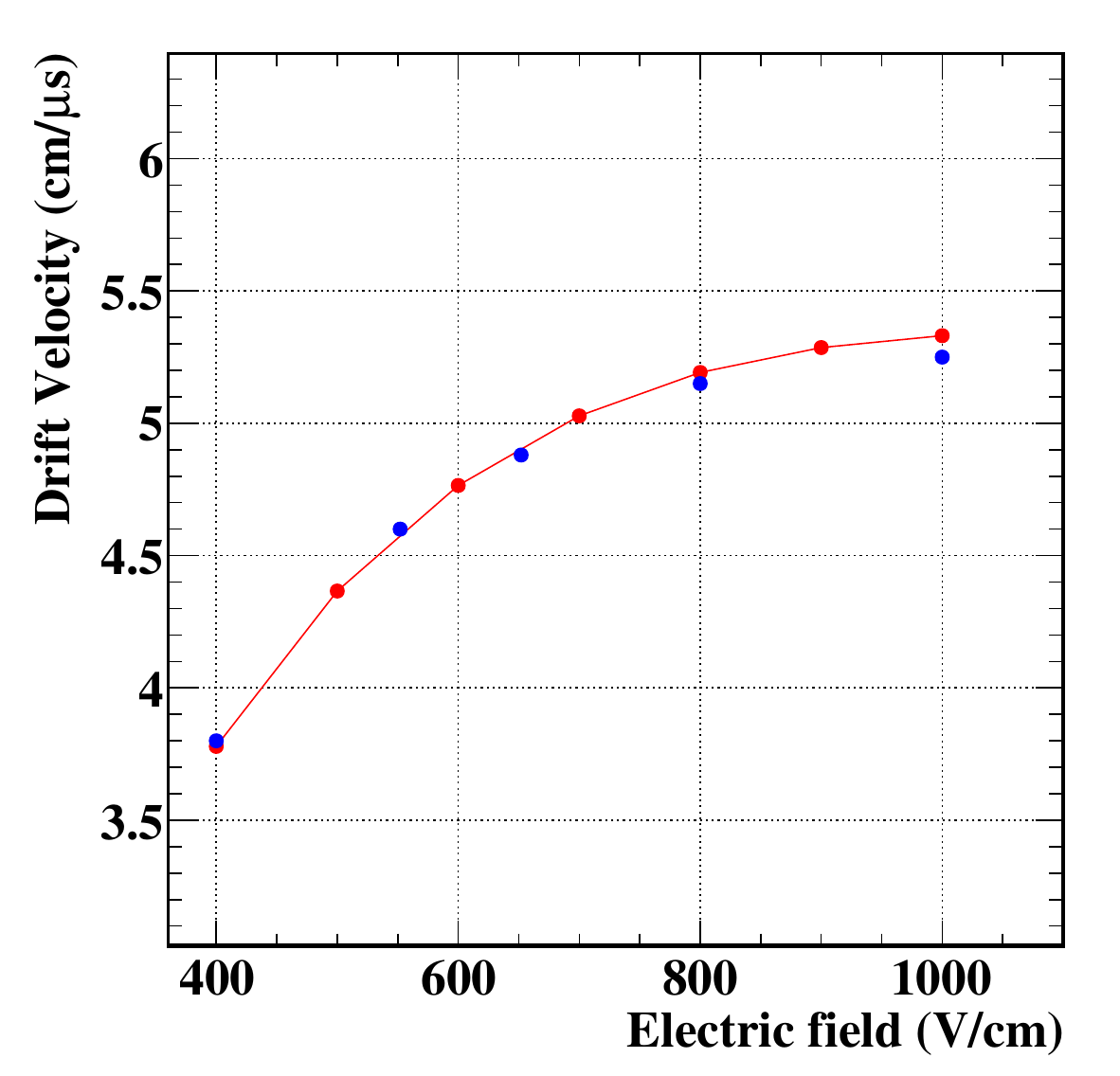}
\caption{Comparison between the measured and the Garfield++ computed drift velocity as a function of the electric field for a gas mixture of Ar:CH4 90:10 + \SI{4}{\percent} Isobutane + O$_2$ (200 ppm). Experimental data extracted from \cite{Huk1987} paper. \label{fig:hukdriftcomp}}
\end{figure}

\begin{figure}[htbp]
\centering
\includegraphics[width=0.49\textwidth]{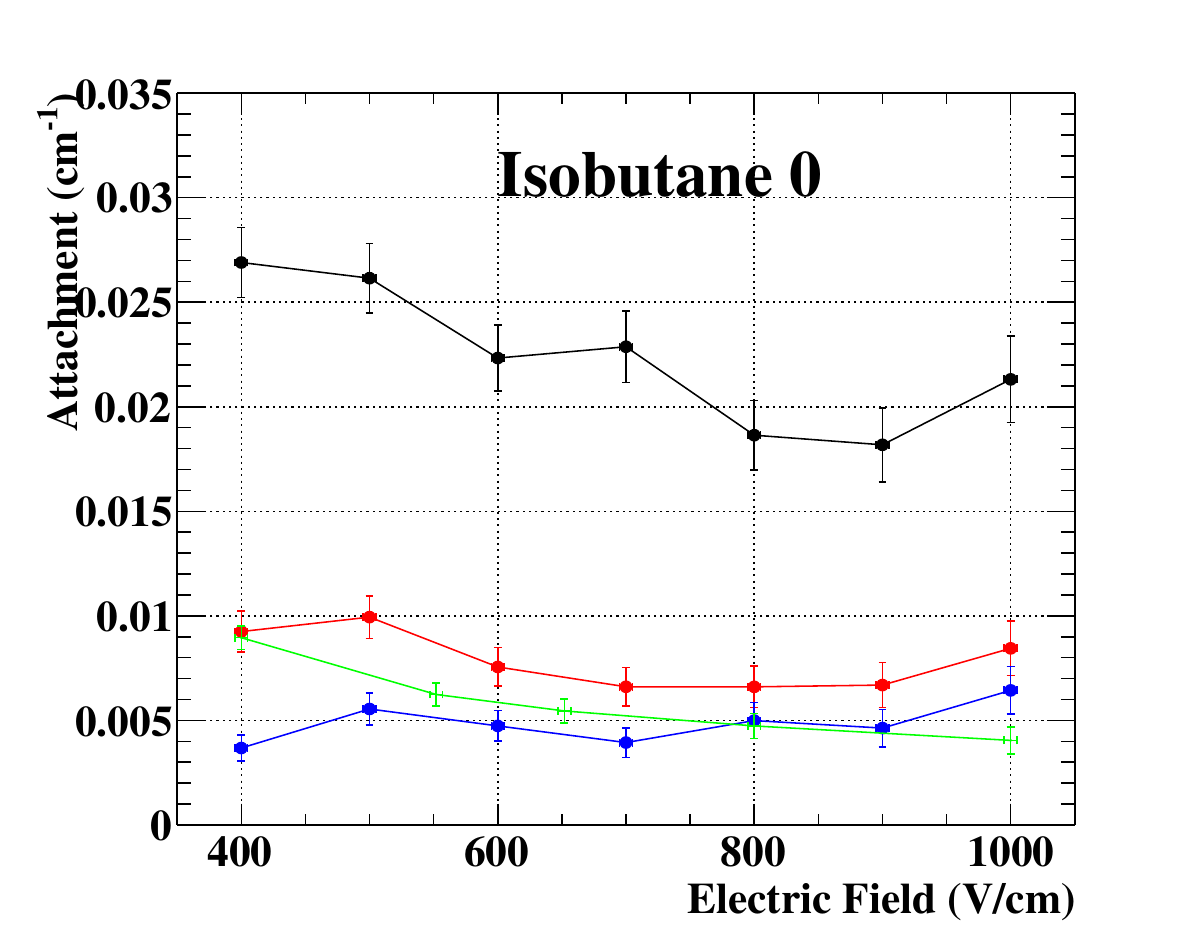}
\includegraphics[width=0.49\textwidth]{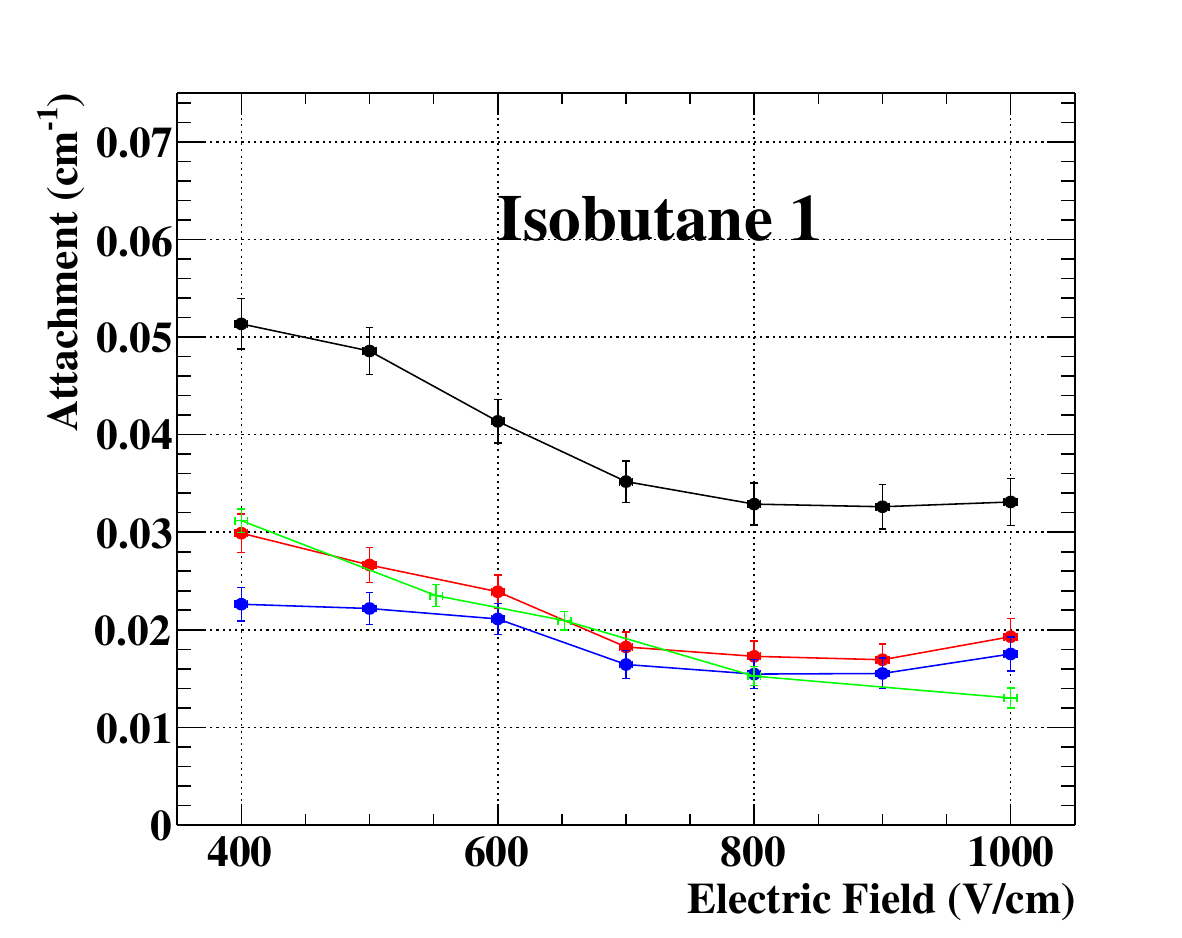}
\includegraphics[width=0.49\textwidth]{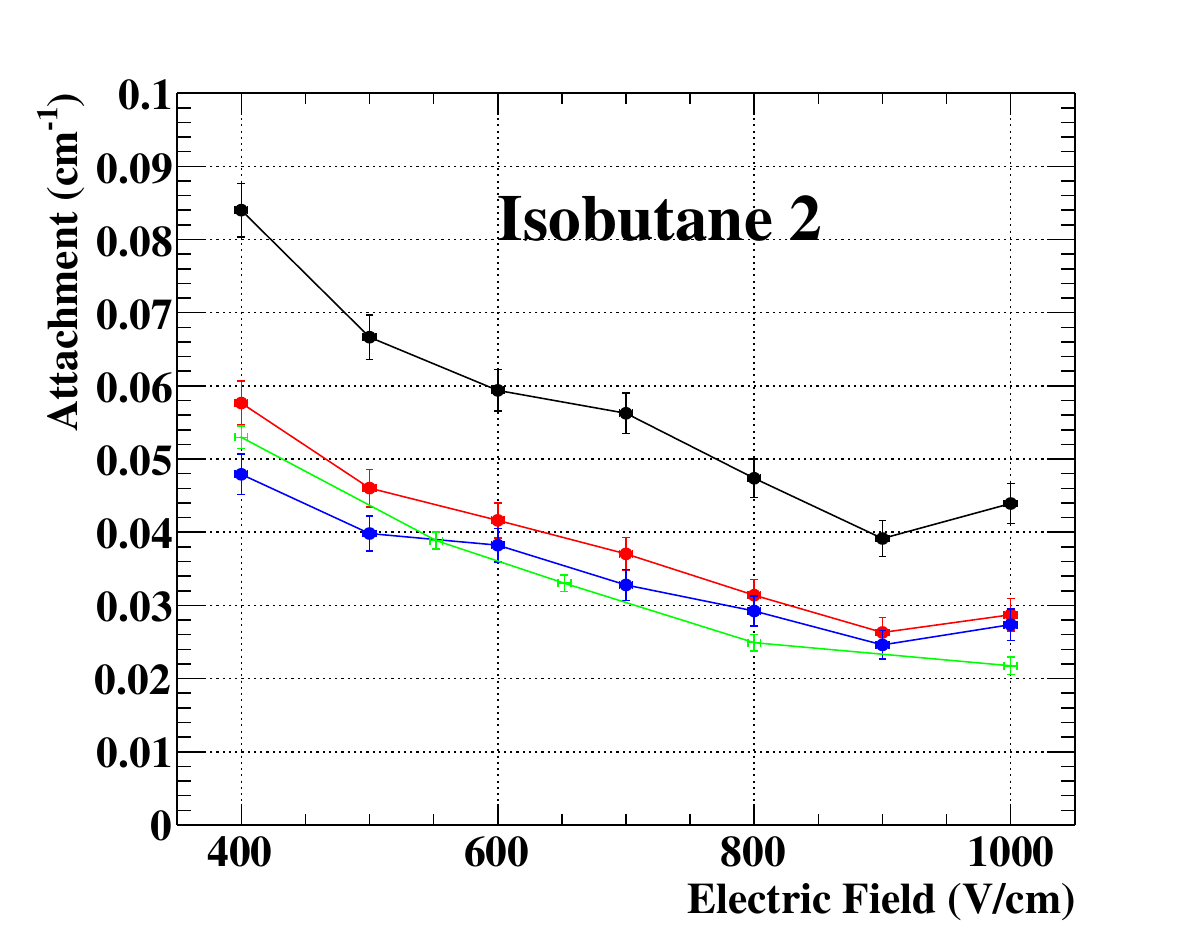}
\includegraphics[width=0.49\textwidth]{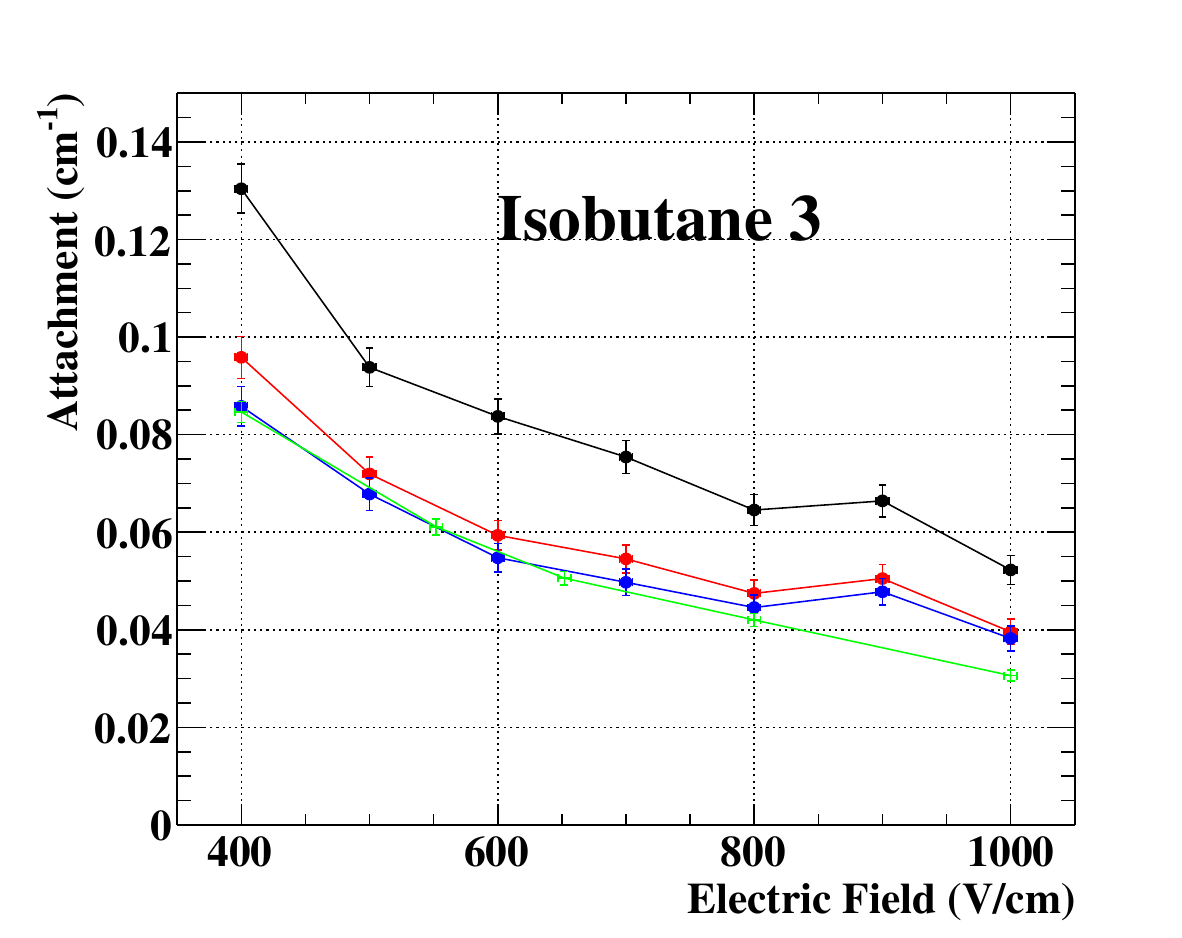}
\includegraphics[width=0.49\textwidth]{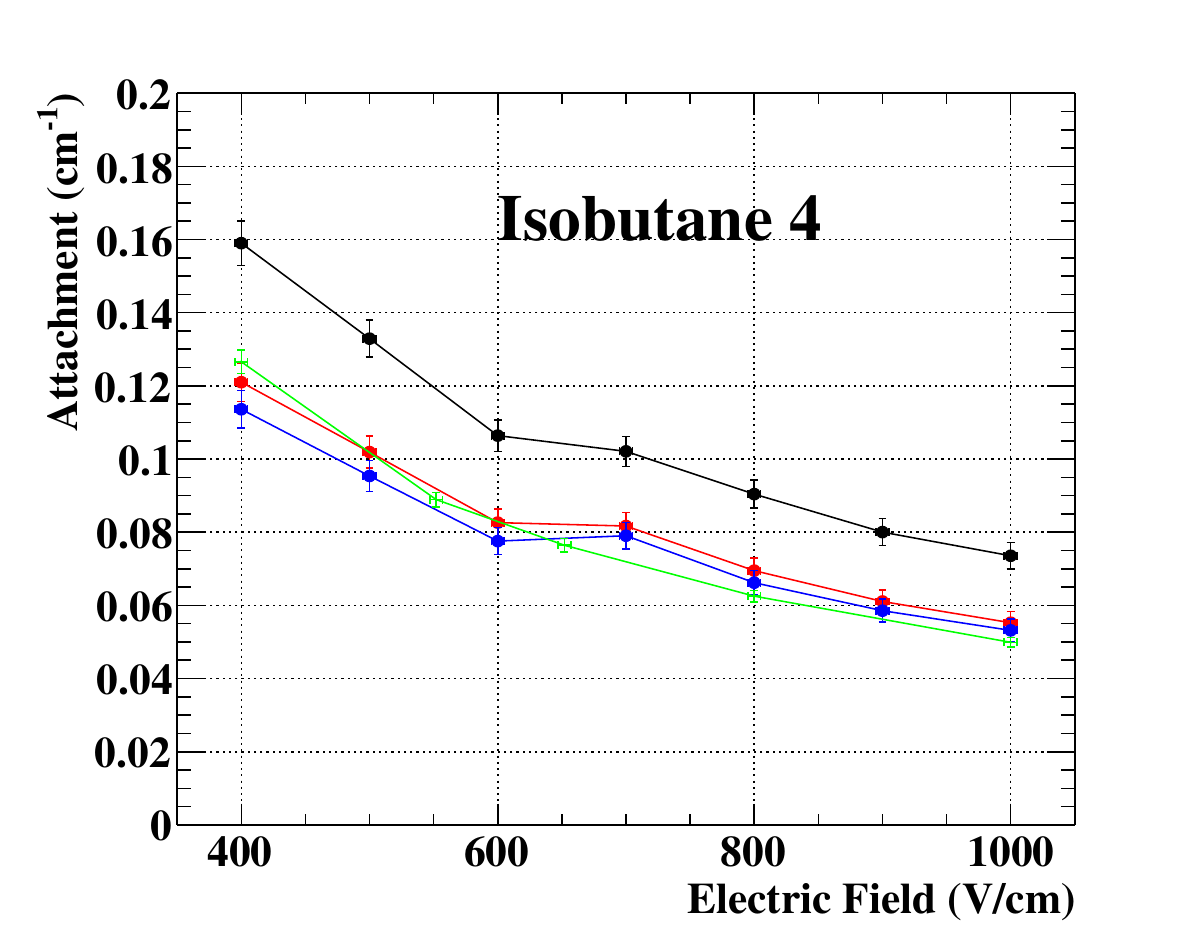}
\hspace{-0.1cm}
\includegraphics[width=0.49\textwidth]{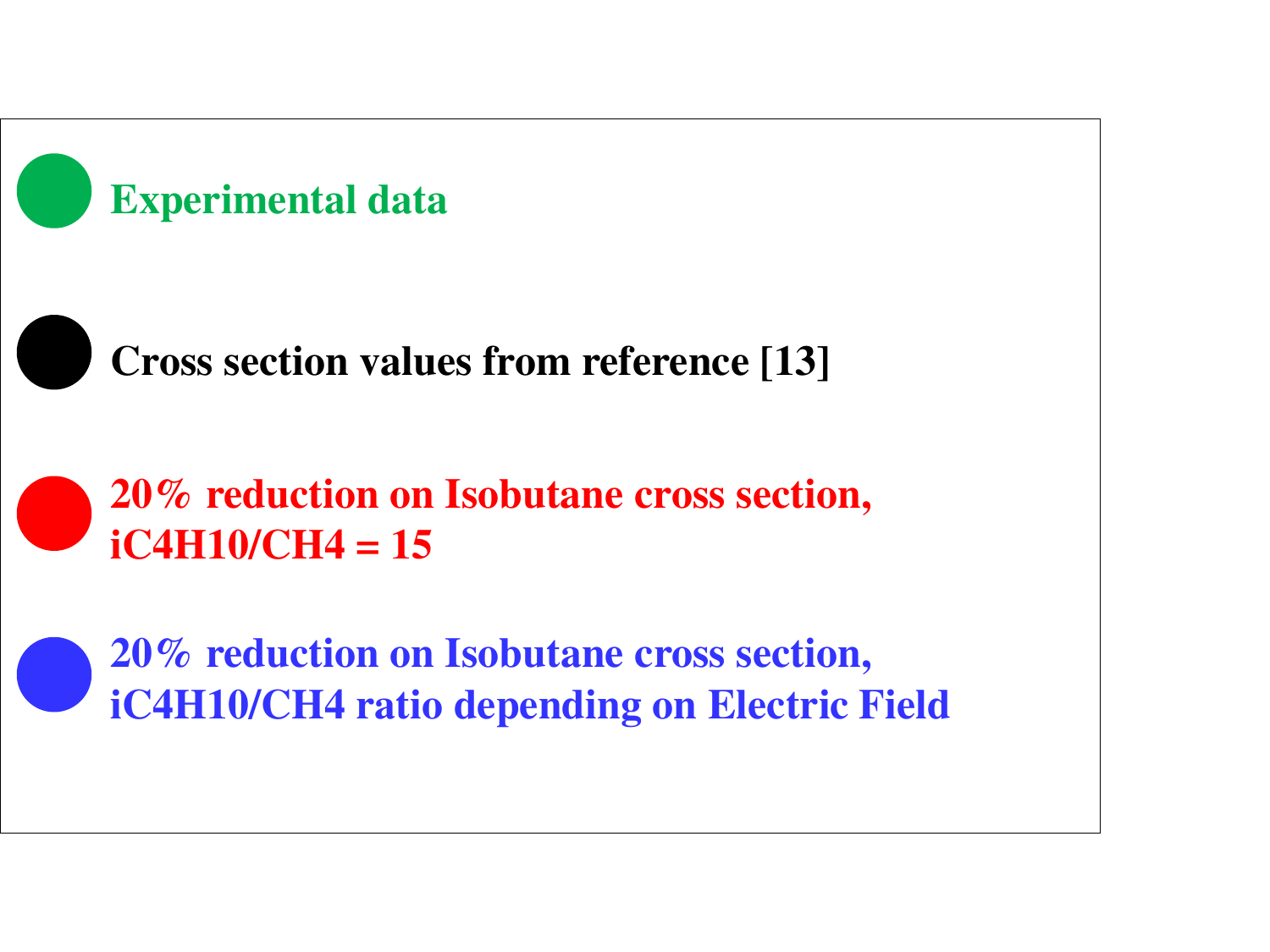}
\caption{Comparison between the attachment 
coefficient measured in paper \cite{Huk1987} and the Garfield++ predictions. The gas mixture is Ar (\SI{90}{\percent}) + CH4 (\SI{10}{\percent}) + O$_2$ (200 ppm) + five different percentages of Isobutane (0, 1, 2, 3, \SI{4}{\percent}, from top to bottom, from left to right). The measurements are shown in green, while the other curves  correspond to Garfield++ simulations based either solely on the numbers from \cite{Shimamori1977} (black) or taking into account the cited \SI{20}{\percent} reduction of Isobutane contribution and the estimates of the contributions to attachment from \cite{Huk1987},  fixing iC4H10/CH4 ratio to 15 (red), or including the measured dependence of this ratio on the electric field (blue). \label{fig:hukcomp}}
\end{figure}

As an additional check on the assumption made for the Isobutane attachment coefficient (the 20\% reduction), in addition to the proper treatment of the three-body process for Oxygen, we show in Fig.~\ref{fig:hukdriftcomp} the comparison of drift velocity as a function of electric field measured in an Ar:CH4 90:10 + \SI{4}{\percent}Isobutane + O$_2$ (200 ppm) mixture \cite{Huk1987} compared with Garfield++ simulations.
We also show, in Fig.~\ref{fig:hukcomp} the comparison  between the measurements (in green) and Garfield++ predictions for the attachment coefficient for the mixture of Ar (\SI{90}{\percent}) + CH4 (\SI{10}{\percent}) + Isobutane (0, 1, 2, 3, \SI{4}{\percent}) + O$_2$ (200 ppm). 
The Garfield++ predictions are based either solely on the numbers from \cite{Shimamori1977} (black curve) or taking into account the cited \SI{20}{\percent} reduction of Isobutane contribution and the estimates of the contributions to attachment from \cite{Huk1987}. In these figures, the red curve uses a fixed i-C$_4$H$_{10}$/CH$_{4}$ ratio of 15, while the blue one includes the measured dependence of this ratio on the electric field.

It should be noted that the assumption made for the comparison with the MEG II measurements of a \SI{20}{\percent} reduction in the de-excitation cross-section for Isobutane is supported by these comparisons too.

\section{Conclusions}

The introduction of Oxygen (along with Isopropanol) into the MEG II drift chamber has resolved localized discharge effects, presumably due to the Malter effect, and stabilized its operation. However, the attachment of electrons to Oxygen is expected to reduce both the gain in the avalanche process and the efficiency in transporting electrons to the anode wires. 

The relative decrease in gain is not a major issue, as it can be compensated by a slight increase in high voltage. Concerning the electron transport efficiency, our measurements confirm, on a more quantitative basis, the information obtained from the reconstruction of hits in the MEG II drift chamber. The results are also in agreement with the Garfield++ simulation when appropriately accounting, as described above, for the various components of the mixture in the Garfield++ code.
Finally, the good resolutions of the detector demonstrate that the attachment due to the presence of Oxygen in a few permille concentration is tolerable with drift distances up to several millimeters.

These findings may serve as a reference for utilizing oxygen as a reliable and safe additive to enhance the stability of current and future drift chambers.

\section{Acknowledgements}

We would like to thank Stephen Biagi and Giuseppe Antonelli for helping us in understanding how to modify Garfield++ to achieve a good comparison with experimental measurements.
Propedeutical measurements were performed at Laboratori Nazionali di Frascati (LNF) inside the laboratories of the CMS group. We would like to thank the LNF Research Division for their hospitality, and especially Simona Giovannella, Luigi Benussi, Alessandro Russo, and the entire technical team for their invaluable assistance.




\end{document}